\documentclass[%
  twocolumn,
 showpacs,
 showkeys,
 preprintnumbers,
 amsmath,amssymb,
 aps,
  pra,
  longbibliography,
 ]{revtex4-1}

\usepackage[dvipsnames]{xcolor}

\usepackage{mathptmx}

\usepackage{amssymb,amsthm,amsmath}

\usepackage{tikz}
\usetikzlibrary{calc}

\usepackage[breaklinks=true,colorlinks=true,anchorcolor=blue,citecolor=blue,filecolor=blue,menucolor=blue,pagecolor=blue,urlcolor=blue,linkcolor=blue]{hyperref}
\usepackage{graphicx}
\usepackage{url}

\usepackage{xcolor}

\begin{document}

\title{New forms of quantum value indefiniteness suggest that incompatible views on contexts are epistemic}


\author{Karl Svozil}
\email{svozil@tuwien.ac.at}
\homepage{http://tph.tuwien.ac.at/~svozil}

\affiliation{Institute for Theoretical Physics,
Vienna  University of Technology,
Wiedner Hauptstrasse 8-10/136,
1040 Vienna,  Austria}

\date{\today}

\begin{abstract}
Extensions of the Kochen-Specker theorem use quantum logics whose classical interpretation suggests a true-implies-value indefiniteness property. This can be interpreted as an indication that any view of a quantum state beyond a single context is epistemic.
\end{abstract}


\keywords{Quantum mechanics, Gleason theorem, Kochen-Specker theorem, Born rule}

\maketitle

\section{Quantum contexts as views on states}
\label{2018-whycontexts-sec1}

Contexts arise naturally in quantum mechanics: they correspond to greatest classical subdomains within the expanse of conceivable quantum propositions.
For all empirical matters, every observable within a particular fixed context can be assumed classical with respect and relative to that context.
Therefore, according to Gleason~\cite{Gleason}, it appears prudent to assume that classical probabilities should be applicable
to such classical mini-universes; and, in particular, when considering observables within a given context.
Gleason formalized this in terms of {\em frame functions} and proceeded to show how the quantum probabilities -- in particular, the Born rule -- can be ``stitched together'' from these classical bits and pieces.
This paper can be seen as a {\it prolegomenon} to this approach; and as a contribution to the ongoing search for its semantics.

Formally, the concept of context can be exposed in two ways:
one is in terms of ``largest possible'' sets of orthogonal pure states; that is, in terms of  (unit) vectors and their linear spans.
Another one is by maximal operators and the perpendicular projection operators in their non-degenerate spectral decomposition.

Let us start by supposing that contexts can be represented by orthonormal bases of Hilbert space.
Due to the spectral theorem this immediately gives rise to an equivalent conception of context:
that as a maximal observable which is formed by some (non-degenerate) spectral sum of the mutually orthogonal perpendicular projection
operators corresponding to the basis states.
This is just the expression of the dual role of perpendicular projection operators in quantum mechanics:
they represent both pure states as well as observable bits; that is, elementary yes-no propositions.

For the sake of an elementary example, suppose one is dealing with (lossless) electron spin state (or photon polarization) measurements.
As there are two outcomes, the associated Hilbert space is two-dimensional.
The two outcomes can be identified with two arbitrary orthogonal normalized vectors therein, forming an orthonormal basis.
Suppose, for the sake of further simplicity, that we parametrize this basis to be the standard Cartesian basis in two-dimensional Hilbert space,
its two vectors being~\cite[Eq.~(1.8)]{mermin-07}
$\vert 0 \rangle = \begin{pmatrix} 1,0 \end{pmatrix}^\intercal$ and
$\vert 1 \rangle = \begin{pmatrix}0,1\end{pmatrix}^\intercal$,
where the superscript symbol ``$\intercal$'' indicates transposition.
Their dyadic products
$\textsf{\textbf{E}}_0
=\vert 0 \rangle \langle 0 \vert
= \begin{pmatrix}1,0\end{pmatrix}^\intercal \otimes \begin{pmatrix}1,0\end{pmatrix}
 = \begin{pmatrix}1&0\\0&0\end{pmatrix}
,
\textsf{\textbf{E}}_1
= \begin{pmatrix}0&0\\0&1\end{pmatrix}
$
form the corresponding (mutually) orthogonal perpendicular projection operators.
These contexts can be either represented in terms of vectors,
like ${\cal C} =\left\{ \vert 0 \rangle , \vert 1 \rangle \right\}$,
or in terms of perpendicular projection operators,
like ${\cal C} =\left\{ \textsf{\textbf{E}}_0  , \textsf{\textbf{E}}_1  \right\}$.

Any two distinct numbers $\lambda_0 \neq \lambda_1$ define a maximal operator
through the ``weighted'' spectral sum
\begin{equation}
\textsf{\textbf{A}} = \lambda_0 \textsf{\textbf{E}}_0  + \lambda_1 \textsf{\textbf{E}}_1
= \lambda_0 \vert 0 \rangle \langle 0 \vert + \lambda_1 \vert 1 \rangle \langle 1 \vert
= \begin{pmatrix}\lambda_0 & 0\\0 & \lambda_1\end{pmatrix}.
\end{equation}
The term ``maximal'' refers to the fact that $\textsf{\textbf{A}}$ ``spans'' a ``classical sub-universe''
of mutually commuting operators through variations of  $f(\textsf{\textbf{A}})
= f(\lambda_0) \textsf{\textbf{E}}_0  + f(\lambda_1) \textsf{\textbf{E}}_1 $,
where $f:\mathbb{R} \mapsto \mathbb{R} $
represents some real valued polynomial or function of a single real argument~\cite[\S~84, Theorems~1{\&}2,p~171]{halmos-vs}.
In particular, this
includes the context ${\cal C} =\left\{ \textsf{\textbf{E}}_0  , \textsf{\textbf{E}}_1  \right\}$
through the two binary functions $f_i(\lambda_j) = \delta_{ij}$, with $i,j \in \{0,1\}$.

\section{Probabilities on contexts in quantum mechanics}
\label{2018-whycontexts-sec2}
Let us concentrate on probabilities next.
As already mentioned,
Gleason~\cite{Gleason} observed that classical observables should obey classical probabilities --
this should be the same for Bayesian and frequentist approaches.
Can we, therefore, hope for the existence of some {\it ``Realding''}
--
that is, some global ontology, some enlarged panorama of ``real physical properties''
--
behind these stitched probabilities?
As it turns out, relative to reasonable assumptions and the absence of exotic options, this is futile.

Formally this issue can be rephrased by recalling that the main formal entities of quantum mechanics are all based on Hilbert space;
that is, on vectors, as well as their relative position and permutations.
A pure state represented as a vector $\vert \psi \rangle$ can be conveniently parameterized or encoded by coordinates referring to the respective bases.
Because of their convenience one chooses orthonormal bases -- that is, contexts -- for such a parametrization.
Why is convenience important?
Because, as has been noted earlier, in finite dimensions $D$
any such context  ${\cal C} \equiv \{\vert {\bf e}_1 \rangle , \vert {\bf e}_2 \rangle, \ldots   ,\vert {\bf e}_D \rangle\}$
can also be interpreted as a maximal set  of co-measurable propositions
${\cal C} \equiv \{\textsf{\textbf{E}}_1   , \textsf{\textbf{E}}_2, \ldots   ,\textsf{\textbf{E}}_D   \}$
with $\textsf{\textbf{E}}_i=\vert {\bf e}_i \rangle \langle {\bf e}_i \vert $, $1 \le i \le D$,
as the latter refers to a complete system of orthogonal perpendicular projections which
are a resolution of the identity operator  $\mathbb{I}_D = \sum_{i=1}^D \textsf{\textbf{E}}_i$.
For any such context, classical Kolmogorov probability theory requires the probabilities $P$ to satisfy the following axioms:
\begin{itemize}
\item[{\bf A1}] -- probabilities are real-valued and non-negative:
$P(\textsf{\textbf{E}}_i) \in \mathbb{R}$,
and
$P(\textsf{\textbf{E}}_i) \ge 0$
for all
$\textsf{\textbf{E}}_i \in {\cal C}$,
or, equivalently,
$1 \le i \le D$;

\item[{\bf A2}] -- probabilities of mutually exclusive observables within contexts are additive:
$
P\left(\sum_{i=1}^{k\le D} \textsf{\textbf{E}}_i\right) = \sum_{i=1}^{k\le D} P\left(\textsf{\textbf{E}}_i\right)
;
$
\item[{\bf A3}] -- probabilities within one context add up to one:
$
P(\mathbb{I}_D) = P\left(\sum_{i=1}^D \textsf{\textbf{E}}_i\right) = 1$.
\end{itemize}

\begin{figure}
\begin{center}
\begin{tikzpicture}  [scale=0.33]



\draw[line width=2pt,
draw=LimeGreen,->] (0,-10cm)
-- (0,10cm)
node[above] {$\vert {\bf e}_2  \rangle$};
\draw[line width=2pt,draw=LimeGreen,->] (-10,0cm)
-- (10cm,0)
node[right] {$\vert {\bf e}_1  \rangle$};


\draw [draw=black!50,dotted]  (0,0) circle[radius=10cm] node[above right] { };
\draw  ({-10/1.3},{-10/1.3}) node[color=black!50,shift={(1.35,0)}] {$r=1$};


\draw [line width=2pt,color=Maroon,->] (-40/5,-30/5) -- (40/5,30/5)  node[above right] {$\vert \psi \rangle$};
\draw [line width=0.5pt,color=Maroon] (40/5,0)  node[color=LimeGreen,font=\scriptsize,shift={(-0.25,-0.25)}] {$\textsf{\textbf{E}}_1\vert \psi \rangle$} -- (40/5,30/5);
\draw [line width=0.5pt,color=Maroon] (0,30/5) node[below left,color=LimeGreen,font=\scriptsize] {$\textsf{\textbf{E}}_2\vert \psi \rangle$} -- (40/5,30/5);
\draw [line width=2pt,color=LimeGreen]  (40/5,0) circle[radius=7pt];
\draw [line width=2pt,color=Maroon]  (40/5,0) circle[radius=2pt];
\draw [line width=2pt,color=LimeGreen]  (0,30/5) circle[radius=7pt];
\draw [line width=2pt,color=Maroon]  (0,30/5) circle[radius=2pt];


 \draw [line width=2pt,color=Maroon,dashed,->] (30/5,-40/5) -- (-30/5,40/5)  node[above left] {$\vert {\bf \varphi} \rangle$};
 \draw [line width=0.5pt,color=Maroon,dashed] (-30/5,0)  node[below,color=LimeGreen,font=\scriptsize] {$\textsf{\textbf{E}}_1\vert \varphi \rangle$} -- (-30/5,40/5);
 \draw [line width=0.5pt,color=Maroon,dashed] (0,40/5) node[color=LimeGreen,font=\scriptsize,shift={(0.95,0)}] {$\textsf{\textbf{E}}_2\vert \varphi \rangle$}  -- (-30/5,40/5);
\draw [line width=2pt,color=LimeGreen]  (-30/5,0) circle[radius=7pt];
\draw [line width=2pt,color=Maroon]  (-30/5,0) circle[radius=2pt];
\draw [line width=2pt,color=LimeGreen]  (0,40/5) circle[radius=7pt];
\draw [line width=2pt,color=Maroon]  (0,40/5) circle[radius=2pt];

\draw [line width=0.5pt,color=LimeGreen] (10,0)  -- ({(40/5)*(20/25)},{(30/5)*(20/25)}) node[color=Maroon,font=\tiny,shift={(-0.42,0.1)}] {$\textsf{\textbf{E}}_\psi \vert {\bf e}_1 \rangle$};
\draw [line width=0.5pt,color=LimeGreen] (10,0)  -- ({(-30/5)*(-3/5)},{(40/5)*(-3/5)}) node[color=Maroon,font=\scriptsize,shift={(0.65,0)}] {$\textsf{\textbf{E}}_\varphi \vert {\bf e}_1 \rangle$};
\draw [line width=2pt,color=Maroon]  ({(40/5)*(20/25)},{(30/5)*(20/25)}) circle[radius=7pt];
\draw [line width=2pt,color=LimeGreen]  ({(40/5)*(20/25)},{(30/5)*(20/25)}) circle[radius=2pt];
\draw [line width=2pt,color=Maroon]  ({(-30/5)*(-3/5)},{(40/5)*(-3/5)}) circle[radius=7pt];
\draw [line width=2pt,color=LimeGreen]  ({(-30/5)*(-3/5)},{(40/5)*(-3/5)}) circle[radius=2pt];

\draw [line width=0.5pt,color=LimeGreen] (0,10)  -- ({(40/5)*(3/5)},{(30/5)*(3/5)}) node[color=Maroon,font=\scriptsize,shift={(0.35,-0.3)}] {$\textsf{\textbf{E}}_\psi \vert {\bf e}_2 \rangle$};
\draw [line width=0.5pt,color=LimeGreen] (0,10)  -- ({(-30/5)*(20/25)},{(40/5)*(20/25)}) node[color=Maroon,font=\scriptsize,shift={(0.65,0)}] {$\textsf{\textbf{E}}_\varphi \vert {\bf e}_2 \rangle$};
\draw [line width=2pt,color=Maroon]  ({(40/5)*(3/5)},{(30/5)*(3/5)}) circle[radius=7pt];
\draw [line width=2pt,color=LimeGreen]  ({(40/5)*(3/5)},{(30/5)*(3/5)}) circle[radius=2pt];
\draw [line width=2pt,color=Maroon]  ({(-30/5)*(20/25)},{(40/5)*(20/25)})  circle[radius=7pt];
\draw [line width=2pt,color=LimeGreen]   ({(-30/5)*(20/25)},{(40/5)*(20/25)})  circle[radius=2pt];

\end{tikzpicture}
\end{center}
\caption{An orthonormal basis forming a context  {\color{LimeGreen} ${\cal C}=\{\vert {\bf e}_1  \rangle , \vert {\bf e}_2  \rangle \}$}
represents a frame of reference from which a ``view'' on a state $\vert \psi \rangle$ can be obtained.
Formally, if the vectors   $\vert \psi \rangle$ and $\vert \varphi \rangle$ are normalized, such that
$\langle \psi \vert \psi \rangle = \langle \varphi \vert \varphi \rangle =  1$, then
the absolute square of the length (norm) of the projections
 $\textsf{\textbf{E}}_1\vert \psi \rangle = \vert {\bf e}_1\rangle \langle {\bf e}_1\vert \psi \rangle $
and
 $\textsf{\textbf{E}}_2\vert \psi \rangle = \vert {\bf e}_2\rangle \langle {\bf e}_2\vert \psi \rangle $
as well as
 $\textsf{\textbf{E}}_1\vert \varphi \rangle = \vert {\bf e}_1\rangle \langle {\bf e}_1\vert \varphi \rangle $
and
 $\textsf{\textbf{E}}_2\vert \varphi \rangle = \vert {\bf e}_2\rangle \langle {\bf e}_2\vert \varphi \rangle $
add up to one.
Conversely, a second  context  {\color{Maroon} ${\cal C'}=\{\vert \psi  \rangle , \vert \varphi  \rangle \}$}
grants a frame of reference from which a ``view'' on the first context $\cal C$ can be obtained.
}
\label{2018-whycontexts-f1-basis}
\end{figure}
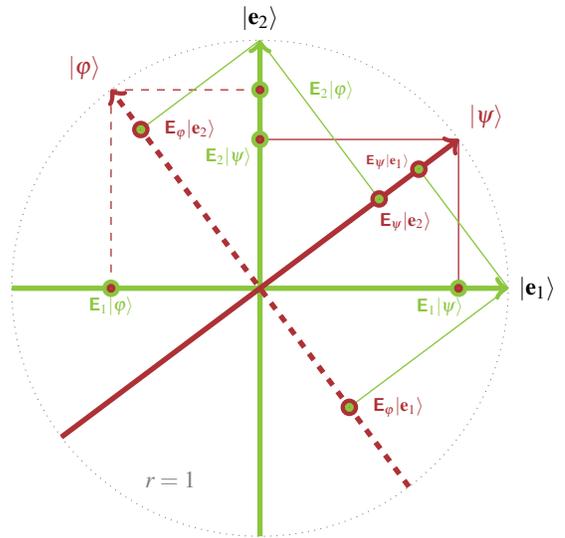

How can probabilities $P_\psi \left( \textsf{\textbf{E}}  \right)$ of propositions formalized by perpendicular projection operators
(or, more generally, observables whose spectral sums contain such propositions)
on given states $\vert \psi \rangle $  be formed which adhere to these axioms?
As already Gleason pointed out in the second paragraph of Ref.~\cite[Sect.~1, p.~885]{Gleason},
there is an {\it ad hoc} way to obtain a probability measure
on Hilbert spaces: a vector $\vert \psi \rangle $
can be ``viewed'' through  a ``probing context''  $\cal C$ as follows:
\begin{itemize}
\item[(i)] For each closed subspace spanned by the vectors $\vert {\bf e}_i \rangle$  in the context $\cal C$, take
the projection   $\textsf{\textbf{E}}_i\vert \psi \rangle$
of $\vert \psi \rangle$ onto  $\vert {\bf e}_i \rangle$.
\item[(ii)] Take the absolute square of the length (norm) of this projection
and identify it with the probability $P_{\psi}\left( \textsf{\textbf{E}}_i  \right)$
of finding the quantum system which is in state  $\vert \psi \rangle $
to be in state $\vert {\bf e}_i \rangle$; that is
(the symbol ``$\dagger$'' stands for the Hermitian adjoint):
\begin{equation}
\begin{split}
P_{\psi}\left( \textsf{\textbf{E}}_i  \right)
=
\left(\textsf{\textbf{E}}_i\vert \psi \rangle\right)^\dagger
\textsf{\textbf{E}}_i\vert \psi \rangle
=
\langle \psi \vert \textsf{\textbf{E}}_i^\dagger
\textsf{\textbf{E}}_i\vert \psi \rangle
\\
= \langle \psi \vert     {\bf e}_i \rangle \underbrace{\langle {\bf e}_i \vert  {\bf e}_i \rangle}_{=1} \langle {\bf e}_i   \vert \psi \rangle
= \langle \psi \vert   {\bf e}_i \rangle  \langle {\bf e}_i \vert    \psi \rangle
= \|  \langle {\bf e}_i  \vert \psi \rangle \|^2
.
\end{split}
\label{2018-whycontexts-glsqnprob}
\end{equation}
\end{itemize}
Because of the mutual orthogonality of the elements in the context $\cal C$, the  Pythagorean theorem
enforces the third axiom {\bf A3} as long all vectors involved are normalized; that is,
has length  (norm)~1.
This situation is depicted in Fig.~\ref{2018-whycontexts-f1-basis}.

The situation is symmetric in a sense which reflects the duality between observable and state observed:
Suppose now that the state $\vert \psi \rangle$ is ``completed''
by other vectors to form an entire context $\cal C'$.
Then one could consider  this context $\cal C'$, including  $\vert \psi \rangle$ to ``probe'' vectors -- now identified as states -- in
the original context  $\cal C$.
Very similarly, probability measures adhering axioms {\bf A1}--{\bf A3} can be constructed by, say,
for instance,  $P_{\textsf{\textbf{E}}_\psi}\left( \textsf{\textbf{E}}_i  \right) $

It is important to keep in mind that, although Gleason's {\it Ansatz} is about a single context $\cal C$
it is valid for all contexts; indeed, formally, for a continuum of contexts represented by the continuum of possible orthonormal bases of $D$-dimensional Hilbert space.
Every such context entails a particular {\em view} on the state   $\vert \psi \rangle $;
and there are a {\em continuum} of such views on the state   $\vert \psi \rangle $.

Furthermore, there is a symmetry between the two contexts
$\cal C$
and
$\cal C'$
involved.
We may call ${\cal C}'$ the ``preparation context'' and ${\cal C}$ the ``measurement context,''
but these denominations are purely conventional.
In this sense, it is a matter of convention if we consider
``$\cal C$ probing $\cal C'$''
or
``$\cal C'$ probing $\cal C$.''

There is one ``privileged view'' on the preparation context $\cal C'$: that is the view obtained if both the preparation and measurement contexts coincide:
${\cal C} = {\cal C}'$. Under such circumstances the observables are value definite: their values coincide with those of the preparation.

\section{Contexts in partition logics and their probabilities}
\label{2018-whycontexts-sec3}

This section is a reminder
rather than an exposition~\cite{svozil-93,dvur-pul-svo,svozil-ql,svozil-2001-eua,svozil-2008-ql,svozil-2016-s,svozil-2016-pu-book} into partition logics.
Suffice it to say that partition logics are probably the most elementary generalization of Boolean algebras:
they are the Boolean subalgebras associated with sets of partitions of a given set which are ``pasted'' or ``stitched'' together at their common elements;
similar to contexts (blocks, subalgebras) in quantum logic.
The main difference is that the latter is a continuous logic based on geometrical entities (vectors), whereas partition logics are discrete,
finite algebraic structures based on sets of partitions of a given set.
Nevertheless, for empirical purposes, it is always possible to come up with a partition logic mimicking the respective quantum logic~\cite{cheval-or}.
Partition logics have two known model realization:  automaton logics~\cite{e-f-moore,schaller-95,schaller-96}
and generalized urn models~\cite{wright,svozil-2005-ln1e,2010-qchocolate}.

Just like classical probabilities on Boolean logics the probabilities on Boolean structures
are formed by a convex summation of all two-valued measures~\cite{svozil-2001-cesena,svozil-2016-s,svozil-2016-pu-book}
-- corresponding to ball types.
Such probabilities will hencefort called {\em (quasi)classical.}

\section{Probabilities on pastings or stitchings of contexts}
\label{2018-whycontexts-sec4}

From dimension $D \ge 3$ onwards, contexts can be non-trivially connected or intertwined~\cite{Gleason}
in up to $D-2$ common elements.
Such intertwining chains of contexts give rise to various apparently ``non-classical'' logics;
and a wealth (some might say a plethora) of publications dealing with ever-increasing ``strange'' or ``magic'' properties of observables
hitherto unheard of in classical physics.
The following logics have a realization in (mostly three-dimensional if not stated otherwise) Hilbert space.
For concrete parametrizations, the reader is either referred to the literature, or to a recent survey~\cite[Chapter~12]{svozil-2016-pu-book}.

On such pastings of contexts, (quasi)classical probabilities and their bounds, termed  {\em conditions of possible experience} by Boole~\cite[p.~229]{Boole-62},
can be obtained in three steps~\cite{svozil-2001-cesena,svozil-2008-ql,svozil-2016-s,svozil-2016-pu-book}:
\begin{itemize}
\item[(i)]
Enumerate all truth assignments (or two-$\{0,1\}$-valued measures or states) $v_i$.
\item[(ii)]
The (quasi)classical probabilities are obtained by the formation of the convex sum $\sum_i \lambda_i v_i$ over all such states obtained in (i), with $0 \le \lambda_i \le 1$ and
$\sum_i \lambda_i =1$.
\item[(ii)]
The Bell-type bounds on probabilities and expectations are attained by
bundling these truth assignments into vectors, one per two-valued measure,
with the  coordinates   representing  the respective values of those states on the atoms (propositions, observables) of the logic;
and by subsequently solving the Hull problem for a convex polytope whose vertices are identified with the vectors formed by all
truth assignments~\cite{froissart-81,cirelson,pitowsky-86,Pit-94}.
\end{itemize}

In what follows some such quantum logics will be enumerated whose quantum probabilities co-exist and sometimes violate their (quasi)classical probabilities,
if they exist.
Such violations can be expected to occur quite regularly, as -- although in both cases the probability axioms {\bf A1}--{\bf A3}
are satisfied for mutually compatible observables -- the quantum probabilities are formed  very differently from the (quasi)classical ones;
that is, not by convex sums as in the (quasi)classical case, but
by scalar products among vectors.

\subsection{Triangular and square logics in four dimensions}

For geometric and algebraic
reasons there is no cyclic pasting of three or four contexts in three dimensions,
but in four dimensions this is possible; as depicted in Fig.~\ref{2018-whycontexts-triangle4D}.
The (quasi)classical probabilities  are enumerated in the Appendices~\ref{2018-whycontexts-app-tri}
and~\ref{2018-whycontexts-app-square}.
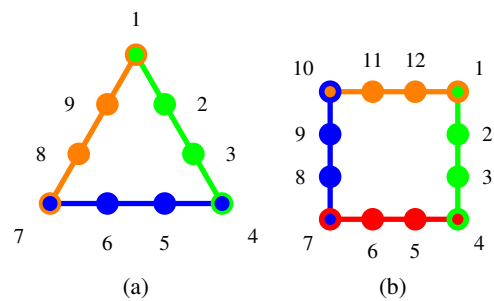
\begin{figure}
\begin{center}
\begin{tabular}{cc}
\begin{tikzpicture}  [scale=0.22]

\newdimen\ms
\ms=0.1cm

\tikzstyle{c3}=[circle,inner sep={\ms/8},minimum size=3*\ms]
\tikzstyle{c2}=[circle,inner sep={\ms/8},minimum size=2*\ms]
\tikzstyle{c1}=[circle,inner sep={\ms/8},minimum size=1*\ms]

\tikzstyle{every path}=[line width=2pt]

\newdimen\R
\R=6cm     



\path
  ({90 + 0 * 360 /3}:\R      ) coordinate(1)
  ({90 + 1 * 360 /3}:\R   ) coordinate(4)
  ({90 + 2 * 360 /3}:\R  ) coordinate(7)
;



\draw [color=orange] (1) -- (4);
\draw [color=blue] (4) -- (7);
\draw [color=green] (7) -- (1);

%
%

\draw (1) coordinate[c3,fill=orange];   %
\draw (1) coordinate[c2,fill=green,label=90:\colorbox{white}{\footnotesize $1$}];  %
%
\node[c3,fill=orange,label={left:\colorbox{white}{\footnotesize  $9$}}] at ( $ (1)!{1/3}!(4) $ ) (2) {};    %
\node[c3,fill=orange,label={left:\colorbox{white}{\footnotesize $8$}}] at ( $ (1)!{2/3}!(4) $ ) (3) {}; %
\draw (4) coordinate[c3,fill=orange];  %
\draw (4) coordinate[c2,fill=blue,label=below left:{\footnotesize \colorbox{white}{$7$}}];  %
%
\node[c3,fill=blue,label={below:{\footnotesize \colorbox{white}{$6$}}}] at ( $ (4)!{1/3}!(7) $ ) (5) {};  %
\node[c3,fill=blue,label={below:{\footnotesize \colorbox{white}{$5$}}}] at ( $ (4)!{2/3}!(7) $ ) (6) {};
\draw (7) coordinate[c3,fill=green];  %
\draw (7) coordinate[c2,fill=blue,label=below right:{\footnotesize  \colorbox{white}{$4$}}];  %
%
\node[c3,fill=green,label={right:\colorbox{white}{\footnotesize $3$}}] at ( $ (7)!{1/3}!(1) $ ) (8) {};  %
\node[c3,fill=green,label={right:\colorbox{white}{\footnotesize $2$}}] at ( $ (7)!{2/3}!(1) $ ) (9) {};
\end{tikzpicture}
&
\begin{tikzpicture}  [scale=0.2]

\newdimen\ms
\ms=0.1cm

\tikzstyle{every path}=[line width=2pt]

\tikzstyle{c3}=[circle,inner sep={\ms/8},minimum size=3*\ms]
\tikzstyle{c2}=[circle,inner sep={\ms/8},minimum size=1.5*\ms]
\tikzstyle{c1}=[circle,inner sep={\ms/8},minimum size=1.1*\ms]

\newdimen\R
\R=6cm     



\path
  ({45 + 0 * 360 /4}:\R      ) coordinate(1)
  ({45 + 1 * 360 /4}:\R   ) coordinate(4)
  ({45 + 2 * 360 /4}:\R  ) coordinate(7)
  ({45 + 3 * 360 /4}:\R  ) coordinate(10)
;


\draw [color=orange] (1) -- (4);
\draw [color=blue]  (4) -- (7);
\draw [color=red] (7) -- (10);
\draw [color=green] (10) -- (1);

%
%

\draw (1) coordinate[c3,fill=orange];   %
\draw (1) coordinate[c2,fill=green,label=45:{\footnotesize $1$}];  %
%
\node[c3,fill=orange,label={above:{\footnotesize  $12$}}] at ( $ (1)!{1/3}!(4) $ ) (2) {};    %
\node[c3,fill=orange,label={above:{\footnotesize $11$}}] at ( $ (1)!{2/3}!(4) $ ) (3) {}; %
\draw (4) coordinate[c3,fill=blue];  %
\draw (4) coordinate[c2,fill=orange,label={180-45}:{\footnotesize {$ {10}$}}];  %
%
\node[c3,fill=blue,label={left:{\footnotesize {$9$}}}] at ( $ (4)!{1/3}!(7) $ ) (5) {};  %
\node[c3,fill=blue,label={left:{\footnotesize {$8$}}}] at ( $ (4)!{2/3}!(7) $ ) (6) {};
\draw (7) coordinate[c3,fill=red];  %
\draw (7) coordinate[c2,fill=blue,label={180+45}:{\footnotesize  {$7$}}];  %
%
\node[c3,fill=red,label={below:{\footnotesize $6$}}] at ( $ (7)!{1/3}!(10) $ ) (8) {};  %
\node[c3,fill=red,label={below:{\footnotesize $5$}}] at ( $ (7)!{2/3}!(10) $ ) (9) {};
\draw (10) coordinate[c3,fill=green];  %
\draw (10) coordinate[c2,fill=red,label= {270+45}:{\footnotesize  {$4$}}];  %
\node[c3,fill=green,label={right:{\footnotesize $3$}}] at ( $ (10)!{1/3}!(1) $ ) (11) {};  %
\node[c3,fill=green,label={right:{\footnotesize $2$}}] at ( $ (10)!{2/3}!(1) $ ) (12) {};
\end{tikzpicture}
\\
(a)&(b)
\end{tabular}
\end{center}
\caption{
Informally, Greechie (or, in another wording, orthogonality) diagrams~\cite{greechie:71}
represent contexts by smooth curves such as straight lines or circles.
The atoms are represented by circles.
Two intertwining contexts are represented by  ``broken'' (not smooth) but connected lines.
(a)
Greechie orthogonality diagram of the triangle logic in four dimensions, realized by (from top)
$1: \frac{1}{2} \begin{pmatrix} 1,1,1,1 \end{pmatrix}^\intercal$,
$2: \frac{1}{\sqrt{2}} \begin{pmatrix} 1,0,-1,0 \end{pmatrix}^\intercal$,
$3: \frac{1}{\sqrt{2}} \begin{pmatrix} 0,1,0,-1 \end{pmatrix}^\intercal$,
$4: \frac{1}{2} \begin{pmatrix} -1,1,-1,1 \end{pmatrix}^\intercal$,
$5: \frac{1}{\sqrt{2}} \begin{pmatrix} 0,1,1,0 \end{pmatrix}^\intercal$,
$6: \frac{1}{\sqrt{2}} \begin{pmatrix} 1,0,0,1 \end{pmatrix}^\intercal$,
$7: \frac{1}{2} \begin{pmatrix} 1,1,-1,-1 \end{pmatrix}^\intercal$,
$8: \frac{1}{\sqrt{2}} \begin{pmatrix} 0,0,1,-1 \end{pmatrix}^\intercal$, and
$9: \frac{1}{\sqrt{2}} \begin{pmatrix} 1,-1,0,0 \end{pmatrix}^\intercal$.
(b)
Greechie orthogonality diagram of the square logic in four dimensions, realized by (from top right)
$1:  \begin{pmatrix} 1,0,0,0 \end{pmatrix}^\intercal$,
$2: \frac{1}{\sqrt{2}} \begin{pmatrix} 0,1,0,1 \end{pmatrix}^\intercal$,
$3: \frac{1}{\sqrt{2}} \begin{pmatrix} 0,1,0,-1 \end{pmatrix}^\intercal$,
$4: \begin{pmatrix} 0,0,1,0 \end{pmatrix}^\intercal$,
$5: \frac{1}{\sqrt{2}} \begin{pmatrix} 1,1,0,0 \end{pmatrix}^\intercal$,
$6: \frac{1}{\sqrt{2}} \begin{pmatrix} 1,-1,0,0 \end{pmatrix}^\intercal$,
$7:  \begin{pmatrix} 0,0,0,1 \end{pmatrix}^\intercal$, and
$8: \frac{1}{\sqrt{2}} \begin{pmatrix} 1,0,1,0 \end{pmatrix}^\intercal$,
$9: \frac{1}{\sqrt{2}} \begin{pmatrix} 1,0,-1,0 \end{pmatrix}^\intercal$,
$10: \begin{pmatrix} 0,1,0,0 \end{pmatrix}^\intercal$,
$11: \frac{1}{\sqrt{2}} \begin{pmatrix} 0,0,1,1 \end{pmatrix}^\intercal$,
$12: \frac{1}{\sqrt{2}} \begin{pmatrix} 0,0,1,-1 \end{pmatrix}^\intercal$.
The associated (quasi)classical probabilities are obtained from a convex summation over all truth assignments,
and listed in the Appendices~\ref{2018-whycontexts-app-tri}
and~\ref{2018-whycontexts-app-square}.
}
\label{2018-whycontexts-triangle4D}
\end{figure}

Summation of the (quasi)classical probabilities on the intertwining atoms of the triangle logic yields
$ p_1+p_4+p_7=
\lambda_1 + \lambda_2 + \lambda_7 +
\lambda_{12}  +
\lambda_{13} + \lambda_{14}  \le 1
$.
However, the axioms of probability theory are too restrictive to allow for quantum violations of these probabilities:
after all, these adjacent vertices are mutually orthogonal, and thus are in the same context (augmented with the fourth atom of that context).
Other inequalities, such as
$p_1+p_2=\lambda_1 + \lambda_2
\le p_5+p_6= (\lambda_1 + \lambda_3 + \lambda_4 + \lambda_8 + \lambda_9)
 + (\lambda_2 + \lambda_5 + \lambda_6 + \lambda_{10} + \lambda_{11})$,
compare vertices with the adjacent ``inner'' atoms;
but again, due to the probability axiom {\bf A3}, the quantum probabilities must obey these inequalities as well.

Komei Fukuda's {\tt cddlib} package~\cite{cdd-pck} can be employed for a calculation of
the hull problem, yielding all Bell-type inequalities associated with the convex polytope whose vertices are associated with the
14 or 34 truth assignments (two-valued measures) on the respective triangle and square logics.
It turns out that all of them are expressions of the axioms {\bf A1}--{\bf A3} which are mandatory also for the quantum probabilities within contexts.

\subsection{Pentagon (pentagram) logic}
\label{2018-whycontexts-pentagons}

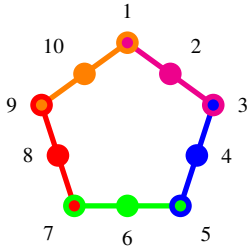
\begin{figure}
\begin{center}
\begin{tikzpicture}  [scale=0.2]

\newdimen\ms
\ms=0.1cm

\tikzstyle{every path}=[line width=2pt]

\tikzstyle{c3}=[circle,inner sep={\ms/8},minimum size=3*\ms]
\tikzstyle{c2}=[circle,inner sep={\ms/8},minimum size=1.5*\ms]
\tikzstyle{c1}=[circle,inner sep={\ms/8},minimum size=1.1*\ms]

\newdimen\R
\R=6cm     



\path
  ({90 + 0 * 360 /5}:\R      ) coordinate(1)
  ({90 + 36 + 0 * 360 /5}:{\R * sqrt((25+10*sqrt(5))/(50+10*sqrt(5)))}      ) coordinate(2)
  ({90 + 1 * 360 /5}:\R   ) coordinate(3)
  ({90 + 36 + 1 * 360 /5}:{\R * sqrt((25+10*sqrt(5))/(50+10*sqrt(5)))}   ) coordinate(4)
  ({90 + 2 * 360 /5}:\R  ) coordinate(5)
  ({90 + 36 + 2 * 360 /5}:{\R * sqrt((25+10*sqrt(5))/(50+10*sqrt(5)))}  ) coordinate(6)
  ({90 + 3 * 360 /5}:\R  ) coordinate(7)
  ({90 + 36 + 3 * 360 /5}:{\R * sqrt((25+10*sqrt(5))/(50+10*sqrt(5)))}  ) coordinate(8)
  ({90 + 4 * 360 /5}:\R     ) coordinate(9)
  ({90 + 36 + 4 * 360 /5}:{\R * sqrt((25+10*sqrt(5))/(50+10*sqrt(5)))}     ) coordinate(10)
;


\draw [color=orange] (1) -- (2) -- (3);
\draw [color=red] (3) -- (4) -- (5);
\draw [color=green] (5) -- (6) -- (7);
\draw [color=blue] (7) -- (8) -- (9);
\draw [color=magenta] (9) -- (10) -- (1);    %

%
%
\draw (1) coordinate[c3,fill=orange,label=90:{\footnotesize $1$}];   %
\draw (1) coordinate[c2,fill=magenta];  %
\draw (2) coordinate[c3,fill=orange,label={above left:\footnotesize $10$}];    %
\draw (3) coordinate[c3,fill=red,label={left:\footnotesize $9$}]; %
\draw (3) coordinate[c2,fill=orange];  %
\draw (4) coordinate[c3,fill=red,label={left:\footnotesize $8$}];  %
\draw (5) coordinate[c3,fill=green,label={below left:\footnotesize $7$}];  %
\draw (5) coordinate[c2,fill=red];  %
\draw (6) coordinate[c3,fill=green,label={below:\footnotesize $6$}];
\draw (7) coordinate[c3,fill=blue,label={below right:\footnotesize $5$}];  %
\draw (7) coordinate[c2,fill=green];  %
\draw (8) coordinate[c3,fill=blue,label={right:\footnotesize $4$}];  %
\draw (9) coordinate[c3,fill=magenta,label={right:\footnotesize $3$}];
\draw (9) coordinate[c2,fill=blue];  %
\draw (10) coordinate[c3,fill=magenta,label={above right:\footnotesize $2$}];  %

\end{tikzpicture}
\end{center}
\caption{Greechie orthogonality diagram of the pentagon (pentagram) logic.
The associated (quasi)classical probabilities are obtained from a convex summation over all truth assignments,
and listed in the Appendix~\ref{2018-whycontexts-app-pent}.
}
\label{2018-whycontexts-pentagon}
\end{figure}

The pentagon (graph theoretically equivalent to a pentagram) logic  is a cyclic stitching or pasting
of five contexts~\cite{wright:pent,kalmbach-83,Beltrametti-1995,Klyachko-2008,Bub-2009,Bub-2010,Badziag-2011}
as depicted in Fig.~\ref{2018-whycontexts-pentagon}.
The (quasi)classical probabilities~\cite[p.~289, Fig.~11.8]{svozil-2016-s}
can be obtained by taking the convex sum of all the 11 two-valued measures~\cite{wright:pent},
as listed in Appendix~\ref{2018-whycontexts-app-pent}.
Because of the convex sum of all $\lambda$'s add up to one, the sum of the (quasi)classical probabilities enumerated in Eq.~(\ref{2018-whycontexts-ppent}), taken merely on the 5 intertwining observables,
yields
\begin{equation}
\begin{aligned}
&p_1+p_3+p_5+p_7+p_9\\
& =
\lambda_1 + \lambda_4 + \lambda_7 +
\lambda_9 + \lambda_{10} + 2\left(
\lambda_2 + \lambda_3 + \lambda_5 +
\lambda_6 + \lambda_8\right)  \\
&\le 2 \sum_{i=1}^{11}\lambda_i  =2
.
\end{aligned}
\end{equation}
This inequality is in violation of quantum predictions~\cite{Bub-2009,Badziag-2011}
of $\sqrt{5}>2$.
Note that, in order to obtain the probabilities on the five intertwining observables (vertices) all of them need to be determined.
However, only adjacent pairs share a common context. So at least three incompatible measurement types are necessary.

\subsection{Specker bug logic with the true--implies--false property}
\label{2018-whycontexts-Specker-bug}

A pasting of two pentagon logics,
the ``Specker bug'' logic, has been introduced~\cite[Fig.~1, p.~182]{kochen2} and used~\cite[$\Gamma_1$, p.~68]{kochen1} by Kochen and Specker
and discussed by many researchers~\cite{redhead,Pitowsky2003395,pitowsky-06}; see also Refs.~\cite[Fig.~B.l. p.~64]{Belinfante-73}, \cite[p.~588-589]{stairs83},
\cite[Sects.~IV, Fig.~2]{clifton-93}  and \cite[p.~39, Fig.~2.4.6]{pulmannova-91}.
It is a pasting~\cite{kalmbach-83,nav:91} of seven contexts in such a tight way
[cf. Fig.~\ref{2018-whycontexts-fbugcombo}(a)]
that preparation of a (quasi)classical system in state
$ {\bf a} $ entails non-occurrence of observable $ {\bf b}$.
As has been observed by Stairs~\cite[p.~588-589]{stairs83} and Clifton~\cite[Sects.~II,III, Fig.~1]{clifton-93,Johansen-1994,Vermaas-1994},
this is no longer the case for quantum states and quantum observables.
Therefore, if one prepares a system in a state  $\vert {\bf a} \rangle$  and measures  $\textsf{\textbf{E}}_b = \vert {\bf b}\rangle \langle {\bf b} \vert$,
associated with state $\vert {\bf b} \rangle$,
then the mere occurrence of $\vert {\bf b} \rangle$ implies the non-classicality of the quantized system.

Again the (quasi)classical probabilities~\cite[p.~286, Fig.~11.5(iii)]{svozil-2016-s} enumerated in Appendix~\ref{2018-whycontexts-app-sbcombo}
can be obtained by taking the convex sum of all the 14 two-valued measures~\cite[p.~579, Table~7]{svozil-2008-ql}.
Pt{\'{a}}k and Pulmannov{\'{a}}~\cite[p.~39, Fig.~2.4.6]{pulmannova-91}
as well as
Pitowsky in Refs.~\cite[p.~402, Fig.~2]{Pitowsky2003395} and~\cite[pp.~224,225, Fig.~10.2]{pitowsky-06}
noted that, for (quasi)classical probabilities -- including ones on partition logics --
the sum of the probabilities on  $\vert {\bf a} \rangle$  and   $\vert {\bf b} \rangle$
must not exceed $\frac{3}{2}$. Therefore both cannot be true at the same time, because this would result in their sum being $2$.
This might be called a {\em true--implies--false property}~\cite{2018-minimalYIYS} ({\it aka}  one-zero rule~\cite{svozil-2006-omni}) on the atoms ${\bf a}$ and ${\bf b}$.

Actually, this classical bound can be tightened by explicity summing the (quasi)classical probabilities
of ${\bf a}$
and ${\bf b}$
enumerated in Eq.~(\ref{2018-whycontexts-pspeckerbug})
Because of the convex sum of all $\lambda$'s add up to one, this yields
yields
\begin{equation}
p_{\bf a}+p_{\bf b}
 =
\lambda_1+\lambda_2+\lambda_3+\lambda_6+\lambda_{13}+\lambda_{14}
\le  \sum_{i=1}^{14}\lambda_i  =1
.
\end{equation}
This inequality is in violation of quantum predictions
for a system prepared in state $\vert {\bf a} \rangle$; in this case~\cite{cabello-1994},  $\frac{10}{9}>1$.

Indeed, Cabello~\cite{cabello-1994} (see also his dissertation~\cite[pp.~55,56]{Cabello-1996-diss}) pointed out that in three dimensions,
 $\vert {\bf a} \rangle$  and   $\vert {\bf b} \rangle$
must be at least an angle
$\angle  ({\bf a},{\bf b}) \ge   \textrm{arcsec} (3) = \textrm{arccos}\left(\frac{1}{3}\right)= \frac{\pi}{2} -  \textrm{arccot}\left(2 \sqrt{2}\right) = \textrm{arctan}\left(2 \sqrt{2}\right)$ apart.
Therefore, the probability to find a state prepared along
$\vert {\bf a} \rangle \equiv \begin{pmatrix} 1,0,0 \end{pmatrix}^\intercal$
in a state
$\vert {\bf b} \rangle \equiv \begin{pmatrix} \cos \angle  ({\bf a},{\bf b}) ,\sin \angle  ({\bf a},{\bf b}) ,0 \end{pmatrix}^\intercal$
cannot exceed $\vert \langle {\bf b} \vert {\bf a} \rangle \vert^2= 1/9$.
Thus in at most one-ninth of all cases will quantum mechanical probabilities violate the classical ones,
as the classical prediction demands zero probability to measure ${\bf b}$, given ${\bf a}$.
(This prediction is relative to the assumption of non-contextuality, such that the truth assignment is independent of the particular context.)
For a concrete ``optimal'' realization~\cite[p.~206, Fig.~1]{tkadlec-96}
(see also~\cite[Fig.~4, p.~5387]{svozil-tkadlec}), take
$\vert {\bf a}\rangle     = \frac{1}{\sqrt{3}}\begin{pmatrix}    1,\sqrt{2},0     \end{pmatrix}^\intercal $ and
$\vert {\bf b}\rangle     = \frac{1}{\sqrt{3}}\begin{pmatrix}   -1,\sqrt{2},0     \end{pmatrix}^\intercal $
which yield $\vert \langle {\bf b} \vert {\bf a}\rangle \vert = \frac{1}{3}$.

%

Another true-implies-false configuration depicted in Fig.~\ref{2018-whycontexts-acsl1}(a)
has an immediate quantum realization~\cite[Table.~1, p.~102201-7]{2015-AnalyticKS} for
$\vert \langle {\bf a} \vert {\bf b}\rangle \vert^2=\frac{1}{2}$, and can be constructively (i.e., algorithmically computable)
extended to arbitrary  angles between non-collinear and non-orthogonal vectors.

\subsection{Combo of Specker bug logic with the true--implies--true as well as inseparability properties}

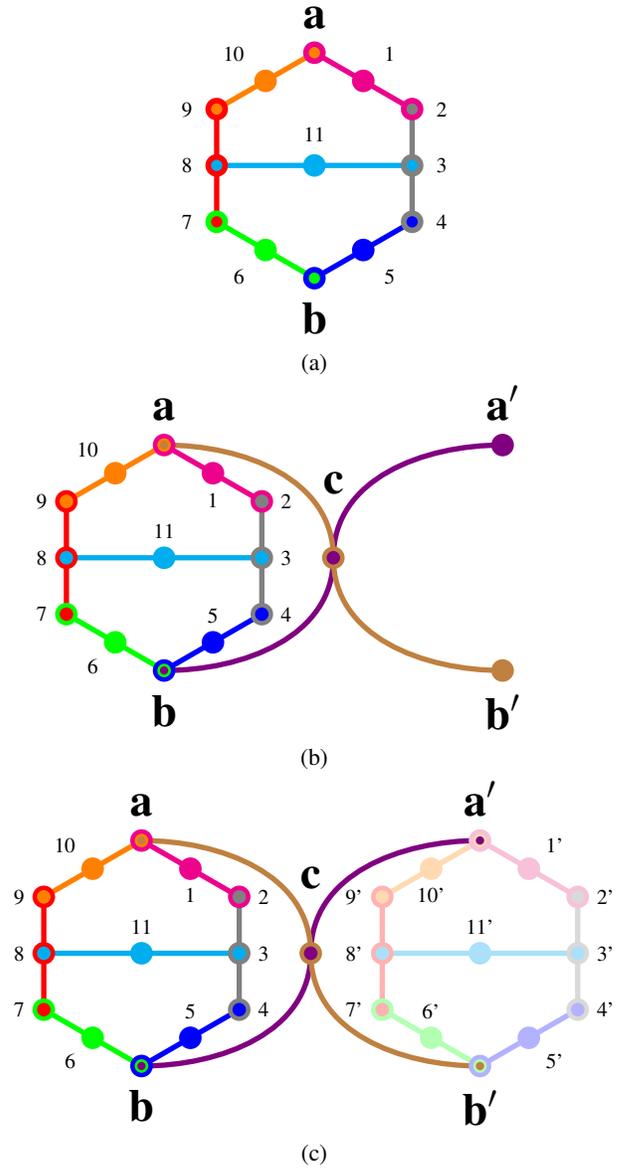
\begin{figure}
\begin{center}
\begin{tabular}{c}
\begin{tikzpicture}  [scale=0.25]

\newdimen\ms
\ms=0.1cm

\tikzstyle{every path}=[line width=2pt]

\tikzstyle{c3}=[circle,inner sep={\ms/8},minimum size=3*\ms]
\tikzstyle{c2}=[circle,inner sep={\ms/8},minimum size=1.5*\ms]
\tikzstyle{c1}=[circle,inner sep={\ms/8},minimum size=1.1*\ms]

\newdimen\R
\R=6cm     



\path
  ({90 + 0 * 360 /6}:\R      ) coordinate(1)
  ({90 + 30 + 0 * 360 /6}:{\R * sqrt(3)/2}      ) coordinate(2)
  ({90 + 1 * 360 /6}:\R   ) coordinate(3)
  ({90 + 30 + 1 * 360 /6}:{\R * sqrt(3)/2}   ) coordinate(4)
  ({90 + 2 * 360 /6}:\R  ) coordinate(5)
  ({90 + 30 + 2 * 360 /6}:{\R * sqrt(3)/2}  ) coordinate(6)
  ({90 + 3 * 360 /6}:\R  ) coordinate(7)
  ({90 + 30 + 3 * 360 /6}:{\R * sqrt(3)/2}  ) coordinate(8)
  ({90 + 4 * 360 /6}:\R     ) coordinate(9)
  ({90 + 30 + 4 * 360 /6}:{\R * sqrt(3)/2}     ) coordinate(10)
  ({90 + 5 * 360 /6}:\R     ) coordinate(11)
  ({90 + 30 + 5 * 360 /6}:{\R * sqrt(3)/2}     ) coordinate(12)
  (0,0) coordinate(13)
;


\draw [color=orange] (1) -- (2) -- (3);
\draw [color=red] (3) -- (4) -- (5);
\draw [color=green] (5) -- (6) -- (7);
\draw [color=blue] (7) -- (8) -- (9);
\draw [color=gray] (9) -- (10) -- (11);    %
\draw [color=magenta] (11) -- (12) -- (1);    %
\draw [color=cyan] (4) -- (10);

%
%

\draw (1) coordinate[c3,fill=magenta,label=90:\LARGE ${\bf a}$];   %
\draw (1) coordinate[c2,fill=orange];  %
\draw (2) coordinate[c3,fill=orange,label=140:{\footnotesize 10}];    %
\draw (3) coordinate[c3,fill=red,label=180:{\footnotesize 9}]; %
\draw (3) coordinate[c2,fill=orange];  %
\draw (4) coordinate[c3,fill=red,label=180:{\footnotesize 8}];  %
\draw (4) coordinate[c2,fill=cyan];  %
\draw (5) coordinate[c3,fill=green,label=180:{\footnotesize 7}];  %
\draw (5) coordinate[c2,fill=red];  %
\draw (6) coordinate[c3,fill=green,label=220:{\footnotesize 6}];
\draw (7) coordinate[c3,fill=blue,label=270:\LARGE ${\bf b}$];  %
\draw (7) coordinate[c2,fill=green];  %
\draw (8) coordinate[c3,fill=blue,label=320:{\footnotesize 5}];  %
\draw (9) coordinate[c3,fill=gray,label=0:{\footnotesize 4}];
\draw (9) coordinate[c2,fill=blue];  %
\draw (10) coordinate[c3,fill=gray,label=0:{\footnotesize 3}];  %
\draw (10) coordinate[c2,fill=cyan];  %
\draw (11) coordinate[c3,fill=magenta,label=0:{\footnotesize 2}];  %
\draw (11) coordinate[c2,fill=gray]; %
\draw (12) coordinate[c3,fill=magenta,label=40:{\footnotesize 1}];  %
\draw (13) coordinate[c3,fill=cyan,label=90:{\footnotesize 11}];  %

\end{tikzpicture}
\\
(a)
\\
\begin{tikzpicture}  [scale=0.25]

\newdimen\ms
\ms=0.1cm

\tikzstyle{every path}=[line width=2pt]

\tikzstyle{c3}=[circle,inner sep={\ms/8},minimum size=3*\ms]
\tikzstyle{c2}=[circle,inner sep={\ms/8},minimum size=1.8*\ms]
\tikzstyle{c1}=[circle,inner sep={\ms/8},minimum size=1.1*\ms]

\newdimen\R
\R=6cm     


\newdimen\K
\K=18cm

\path
  ({90 + 0 * 360 /6}:\R      ) coordinate(1)
  ({90 + 30 + 0 * 360 /6}:{\R * sqrt(3)/2}      ) coordinate(2)
  ({90 + 1 * 360 /6}:\R   ) coordinate(3)
  ({90 + 30 + 1 * 360 /6}:{\R * sqrt(3)/2}   ) coordinate(4)
  ({90 + 2 * 360 /6}:\R  ) coordinate(5)
  ({90 + 30 + 2 * 360 /6}:{\R * sqrt(3)/2}  ) coordinate(6)
  ({90 + 3 * 360 /6}:\R  ) coordinate(7)
  ({90 + 30 + 3 * 360 /6}:{\R * sqrt(3)/2}  ) coordinate(8)
  ({90 + 4 * 360 /6}:\R     ) coordinate(9)
  ({90 + 30 + 4 * 360 /6}:{\R * sqrt(3)/2}     ) coordinate(10)
  ({90 + 5 * 360 /6}:\R     ) coordinate(11)
  ({90 + 30 + 5 * 360 /6}:{\R * sqrt(3)/2}     ) coordinate(12)
  (0,0) coordinate(13)

  ({90 + 0 * 360 /6}:\R     ) + (\K,0)  coordinate(21)
  ({90 + 30 + 0 * 360 /6}:{\R * sqrt(3)/2}     ) + (\K,0)  coordinate(22)
  ({90 + 1 * 360 /6}:\R  ) + (\K,0)  coordinate(23)
  ({90 + 30 + 1 * 360 /6}:{\R * sqrt(3)/2}  ) + (\K,0)  coordinate(24)
  ({90 + 2 * 360 /6}:\R  ) + (\K,0) coordinate(25)
  ({90 + 30 + 2 * 360 /6}:{\R * sqrt(3)/2}  ) + (\K,0)  coordinate(26)
  ({90 + 3 * 360 /6}:\R  ) + (\K,0)  coordinate(27)
  ({90 + 30 + 3 * 360 /6}:{\R * sqrt(3)/2}  ) + (\K,0)  coordinate(28)
  ({90 + 4 * 360 /6}:\R    ) + (\K,0)  coordinate(29)
  ({90 + 30 + 4 * 360 /6}:{\R * sqrt(3)/2}    ) + (\K,0)  coordinate(210)
  ({90 + 5 * 360 /6}:\R    ) + (\K,0)  coordinate(211)
  ({90 + 30 + 5 * 360 /6}:{\R * sqrt(3)/2}    ) + (\K,0)  coordinate(212)
  (0,0) + (\K,0) coordinate(213)

   (0,0) + ({\K/2},0) coordinate(40)
;


\draw [color=orange] (1) -- (2) -- (3);
\draw [color=red] (3) -- (4) -- (5);
\draw [color=green] (5) -- (6) -- (7);
\draw [color=blue] (7) -- (8) -- (9);
\draw [color=gray] (9) -- (10) -- (11);    %
\draw [color=magenta] (11) -- (12) -- (1);    %
\draw [color=cyan] (4) -- (10);

\draw [color=violet] (7)  to   [out=0,in=270] (40) to [out=90,in=180] (21);
\draw [color=brown] (27)  to   [out=180,in=270] (40) to [out=90,in=0] (1);

%
%

\draw (1) coordinate[c3,fill=magenta,label=90:\LARGE ${\bf a}$];   %
\draw (1) coordinate[c2,fill=orange];  %
\draw (1) coordinate[c1,fill=brown];  %
\draw (2) coordinate[c3,fill=orange,label={[label distance=-2]140:{\footnotesize 10}}];    %
\draw (3) coordinate[c3,fill=red,label={[label distance=-2]180:{\footnotesize 9}}]; %
\draw (3) coordinate[c2,fill=orange];  %
\draw (4) coordinate[c3,fill=red,label={[label distance=-2]180:{\footnotesize 8}}];  %
\draw (4) coordinate[c2,fill=cyan];  %
\draw (5) coordinate[c3,fill=green,label={[label distance=-2]180:{\footnotesize 7}}];  %
\draw (5) coordinate[c2,fill=red];  %
\draw (6) coordinate[c3,fill=green,label={[label distance=-2]220:{\footnotesize 6}}];
\draw (7) coordinate[c3,fill=blue,label=270:\LARGE ${\bf b}$];  %
\draw (7) coordinate[c2,fill=green];  %
\draw (7) coordinate[c1,fill=violet];  %
\draw (8) coordinate[c3,fill=blue,label={[label distance=-2]90:{\footnotesize 5}}];  %
\draw (9) coordinate[c3,fill=gray,label={[label distance=-2]0:{\footnotesize 4}}];
\draw (9) coordinate[c2,fill=blue];  %
\draw (10) coordinate[c3,fill=gray,label={[label distance=-2]0:{\footnotesize 3}}];  %
\draw (10) coordinate[c2,fill=cyan];  %
\draw (11) coordinate[c3,fill=magenta,label={[label distance=-2]0:{\footnotesize 2}}];  %
\draw (11) coordinate[c2,fill=gray]; %
\draw (12) coordinate[c3,fill=magenta,label={[label distance=-2]270:{\footnotesize 1}}];  %
\draw (13) coordinate[c3,fill=cyan,label={[label distance=-2]90:{\footnotesize 11}}];  %

\draw (21) coordinate[c3,fill=violet,label=90:\LARGE ${\bf a}'$];   %

\draw (27) coordinate[c3,fill=brown,label=270:\LARGE ${\bf b}'$];  %

\draw (40) coordinate[c3,fill=brown,label={[label distance=15]90:\LARGE ${\bf c}$}];  %
\draw (40) coordinate[c2,fill=violet];  %

\end{tikzpicture}
\\
(b)
\\
\begin{tikzpicture}  [scale=0.25]

\newdimen\ms
\ms=0.1cm

\tikzstyle{every path}=[line width=2pt]

\tikzstyle{c3}=[circle,inner sep={\ms/8},minimum size=3*\ms]
\tikzstyle{c2}=[circle,inner sep={\ms/8},minimum size=1.8*\ms]
\tikzstyle{c1}=[circle,inner sep={\ms/8},minimum size=1.1*\ms]

\newdimen\R
\R=6cm     


\newdimen\K
\K=18cm

\path
  ({90 + 0 * 360 /6}:\R      ) coordinate(1)
  ({90 + 30 + 0 * 360 /6}:{\R * sqrt(3)/2}      ) coordinate(2)
  ({90 + 1 * 360 /6}:\R   ) coordinate(3)
  ({90 + 30 + 1 * 360 /6}:{\R * sqrt(3)/2}   ) coordinate(4)
  ({90 + 2 * 360 /6}:\R  ) coordinate(5)
  ({90 + 30 + 2 * 360 /6}:{\R * sqrt(3)/2}  ) coordinate(6)
  ({90 + 3 * 360 /6}:\R  ) coordinate(7)
  ({90 + 30 + 3 * 360 /6}:{\R * sqrt(3)/2}  ) coordinate(8)
  ({90 + 4 * 360 /6}:\R     ) coordinate(9)
  ({90 + 30 + 4 * 360 /6}:{\R * sqrt(3)/2}     ) coordinate(10)
  ({90 + 5 * 360 /6}:\R     ) coordinate(11)
  ({90 + 30 + 5 * 360 /6}:{\R * sqrt(3)/2}     ) coordinate(12)
  (0,0) coordinate(13)

  ({90 + 0 * 360 /6}:\R     ) + (\K,0)  coordinate(21)
  ({90 + 30 + 0 * 360 /6}:{\R * sqrt(3)/2}     ) + (\K,0)  coordinate(22)
  ({90 + 1 * 360 /6}:\R  ) + (\K,0)  coordinate(23)
  ({90 + 30 + 1 * 360 /6}:{\R * sqrt(3)/2}  ) + (\K,0)  coordinate(24)
  ({90 + 2 * 360 /6}:\R  ) + (\K,0) coordinate(25)
  ({90 + 30 + 2 * 360 /6}:{\R * sqrt(3)/2}  ) + (\K,0)  coordinate(26)
  ({90 + 3 * 360 /6}:\R  ) + (\K,0)  coordinate(27)
  ({90 + 30 + 3 * 360 /6}:{\R * sqrt(3)/2}  ) + (\K,0)  coordinate(28)
  ({90 + 4 * 360 /6}:\R    ) + (\K,0)  coordinate(29)
  ({90 + 30 + 4 * 360 /6}:{\R * sqrt(3)/2}    ) + (\K,0)  coordinate(210)
  ({90 + 5 * 360 /6}:\R    ) + (\K,0)  coordinate(211)
  ({90 + 30 + 5 * 360 /6}:{\R * sqrt(3)/2}    ) + (\K,0)  coordinate(212)
  (0,0) + (\K,0) coordinate(213)

   (0,0) + ({\K/2},0) coordinate(40)
;


\draw [color=orange] (1) -- (2) -- (3);
\draw [color=red] (3) -- (4) -- (5);
\draw [color=green] (5) -- (6) -- (7);
\draw [color=blue] (7) -- (8) -- (9);
\draw [color=gray] (9) -- (10) -- (11);    %
\draw [color=magenta] (11) -- (12) -- (1);    %
\draw [color=cyan] (4) -- (10);

\draw [color=orange!30!white] (21) -- (22) -- (23);
\draw [color=red!30!white] (23) -- (24) -- (25);
\draw [color=green!30!white] (25) -- (26) -- (27);
\draw [color=blue!30!white] (27) -- (28) -- (29);
\draw [color=gray!30!white] (29) -- (210) -- (211);    %
\draw [color=magenta!30!white] (211) -- (212) -- (21);    %
\draw [color=cyan!30!white] (24) -- (210);

\draw [color=violet] (7)  to   [out=0,in=270] (40) to [out=90,in=180] (21);
\draw [color=brown] (27)  to   [out=180,in=270] (40) to [out=90,in=0] (1);

%
%

\draw (1) coordinate[c3,fill=magenta,label=90:\LARGE ${\bf a}$];   %
\draw (1) coordinate[c2,fill=orange];  %
\draw (1) coordinate[c1,fill=brown];  %
\draw (2) coordinate[c3,fill=orange,label={[label distance=-2]140:{\footnotesize 10}}];    %
\draw (3) coordinate[c3,fill=red,label={[label distance=-2]180:{\footnotesize 9}}]; %
\draw (3) coordinate[c2,fill=orange];  %
\draw (4) coordinate[c3,fill=red,label={[label distance=-2]180:{\footnotesize 8}}];  %
\draw (4) coordinate[c2,fill=cyan];  %
\draw (5) coordinate[c3,fill=green,label={[label distance=-2]180:{\footnotesize 7}}];  %
\draw (5) coordinate[c2,fill=red];  %
\draw (6) coordinate[c3,fill=green,label={[label distance=-2]220:{\footnotesize 6}}];
\draw (7) coordinate[c3,fill=blue,label=270:\LARGE ${\bf b}$];  %
\draw (7) coordinate[c2,fill=green];  %
\draw (7) coordinate[c1,fill=violet];  %
\draw (8) coordinate[c3,fill=blue,label={[label distance=-2]90:{\footnotesize 5}}];  %
\draw (9) coordinate[c3,fill=gray,label={[label distance=-2]0:{\footnotesize 4}}];
\draw (9) coordinate[c2,fill=blue];  %
\draw (10) coordinate[c3,fill=gray,label={[label distance=-2]0:{\footnotesize 3}}];  %
\draw (10) coordinate[c2,fill=cyan];  %
\draw (11) coordinate[c3,fill=magenta,label={[label distance=-2]0:{\footnotesize 2}}];  %
\draw (11) coordinate[c2,fill=gray]; %
\draw (12) coordinate[c3,fill=magenta,label={[label distance=-2]270:{\footnotesize 1}}];  %
\draw (13) coordinate[c3,fill=cyan,label={[label distance=-2]90:{\footnotesize 11}}];  %

\draw (21) coordinate[c3,fill=magenta!30!white,label=90:\LARGE ${\bf a}'$];   %
\draw (21) coordinate[c2,fill=orange!30!white];  %
\draw (21) coordinate[c1,fill=violet];  %
\draw (22) coordinate[c3,fill=orange!30!white,label={[label distance=-2]270:{\footnotesize 10'}}];    %
\draw (23) coordinate[c3,fill=red!30!white,label={[label distance=-2]180:{\footnotesize 9'}}]; %
\draw (23) coordinate[c2,fill=orange!30!white];  %
\draw (24) coordinate[c3,fill=red!30!white,label={[label distance=-2]180:{\footnotesize 8'}}];  %
\draw (24) coordinate[c2,fill=cyan!30!white];  %
\draw (25) coordinate[c3,fill=green!30!white,label={[label distance=-2]180:{\footnotesize 7'}}];  %
\draw (25) coordinate[c2,fill=red!30!white];  %
\draw (26) coordinate[c3,fill=green!30!white,label={[label distance=-2]90:{\footnotesize 6'}}];
\draw (27) coordinate[c3,fill=blue!30!white,label=270:\LARGE ${\bf b}'$];  %
\draw (27) coordinate[c2,fill=green!30!white];  %
\draw (27) coordinate[c1,fill=brown];  %
\draw (28) coordinate[c3,fill=blue!30!white,label={[label distance=-2]320:{\footnotesize 5'}}];  %
\draw (29) coordinate[c3,fill=gray!30!white,label={[label distance=-2]0:{\footnotesize 4'}}];
\draw (29) coordinate[c2,fill=blue!30!white];  %
\draw (210) coordinate[c3,fill=gray!30!white,label={[label distance=-2]0:{\footnotesize 3'}}];  %
\draw (210) coordinate[c2,fill=cyan!30!white];  %
\draw (211) coordinate[c3,fill=magenta!30!white,label={[label distance=-2]0:{\footnotesize 2'}}];  %
\draw (211) coordinate[c2,fill=gray!30!white]; %
\draw (212) coordinate[c3,fill=magenta!30!white,label={[label distance=-2]40:{\footnotesize 1'}}];  %
\draw (213) coordinate[c3,fill=cyan!30!white,label={[label distance=-2]90:{\footnotesize 11'}}];  %

\draw (40) coordinate[c3,fill=brown,label={[label distance=15]90:\LARGE ${\bf c}$}];  %
\draw (40) coordinate[c2,fill=violet];  %
\end{tikzpicture}
\\
(c)
\end{tabular}
\end{center}
\caption{Greechie orthogonality diagram of (a) the Specker bug logic~\cite[Fig.~1, p.~182]{kochen2}.
A proof that, if the system is prepared in state ${\bf a}$, then classical (non-contextual) truth assignments require ${\bf b}$ not to occur, proceeds as follows:
In such a truth assignment, as {\it per} axiom {\bf A3}, there is only one true atom per context; all the others have to be false.
In a proof by contradiction, suppose that both ${\bf a}$ and ${\bf b}$ are true.
Then all atoms connected to them (2,4,7,9) must be false.
This in turn requires that the observables (3,8) connecting them must both be true.
Alas, those two observables (3,8) are connected by a ``middle'' context {\color{cyan}$\{3,11,8\}$}.
But the occurrence of two true observables within the same context is forbidden by axiom {\bf A3}.
The only consistent alternative is to disallow ${\bf b}$ to be true if ${\bf a}$ is assumed to be true; or, conversely,
 to disallow ${\bf a}$ to be true if ${\bf b}$ is assumed to be true.
(b)~Greechie orthogonality diagram of a  Specker bug logic  extended by two contexts which has the true-implies-true property on ${\bf a'}$, given ${\bf a}$ to be true~\cite[$\Gamma_1$, p.~68]{kochen1}.
(c)~Greechie orthogonality diagram of a combo of two Specker bug logics~\cite[$\Gamma_3$, p.~70]{kochen1}.
If ${\bf a}$ is assumed to be true
then the remaining atoms in the context {\color{violet} $\left\{ {\bf a}, {\bf c}, {\bf b}' \right\}$} connecting ${\bf a}$ with ${\bf b'}$,
and, in particular, ${\bf c}$, have to be false.
Also if ${\bf a}$ is  true then ${\bf b}$ is false.
Therefore, ${\bf a}'$ needs to be true if ${\bf b}$ and ${\bf c}$ both are false, because they form the context {\color{brown} $\left\{ {\bf b}, {\bf c}, {\bf a}' \right\}$}.
This argument is valid even in the absence of a second Specker bug logic.
Introduction of a second Specker bug logic ensures the converse: whenever ${\bf a}'$ is true, ${\bf a}$ must be true as well.
Therefore ${\bf a}$ and ${\bf a}'$ (and by symmetry also ${\bf b}$ and ${\bf b}'$) cannot be separated by any truth assignment.}
\label{2018-whycontexts-fbugcombo}
\end{figure}

This non-classical behaviour can be ``boosted''
by an extension of the Specker bug logic~\cite[$\Gamma_1$, p.~68]{kochen1}, including two additional contexts
{\color{violet} $\left\{ {\bf a}, {\bf c} , {\bf b}' \right\}$} as well as
{\color{brown} $\left\{ {\bf b}, {\bf c} , {\bf a}' \right\}$},
as depicted in Fig.~\ref{2018-whycontexts-fbugcombo}(b).
It implements a {\em true-implies-true property}~\cite{2018-minimalYIYS} ({\it aka}  one-one rule~\cite{svozil-2006-omni})  for ${\bf a}$ and ${\bf a}'$.
Cabello's bound
on the angle
$\angle ({\bf a},{\bf b})$
between
${\bf a}$
and
${\bf b}$
mentioned earlier
results in bounds between
${\bf a}$
and
${\bf a}'$
as well as
${\bf b}$
and
${\bf b}'$:
since
${\bf a}$
and
${\bf b}'$
as well as
${\bf b}$
and
${\bf a}'$
are orthogonal, that is,
$
\angle ({\bf a},{\bf b}') =
\angle ({\bf b},{\bf a}') = \frac{\pi}{2}
$,
it follows for planar configurations that
$\angle ({\bf a},{\bf a}')
=
\angle ({\bf b},{\bf a}') -  \angle ({\bf a},{\bf b})
\le
 \frac{\pi}{2} - \textrm{arccos} \left(\frac{1}{3} \right)=
\textrm{arccot} \left( 2\sqrt{2} \right)= \textrm{arccsc} \left( {3} \right)=  \textrm{arcsin} \left(\frac{1}{3} \right)
$.
For symmetry reasons, the same estimate holds for planar configurations between
${\bf b}$
and
${\bf b}'$.
For non-planar configurations the angles must be even less than for planar ones.

True-implies-true properties have also been
studied by Stairs~\cite[p.~588-589, note added in proof]{stairs83};
Clifton~\cite[Sects.~II,III, Fig.~1]{clifton-93,Johansen-1994,Vermaas-1994}
presents a similar argument, based upon  another true-implies-true logic
inspired by Bell~\cite[Fig.~C.l. p.~67]{Belinfante-73} (cf. also Pitowsky~\cite[p.~394]{Pitowsky-1982-subs}),
on the Specker bug logic~\cite[Sects.~IV, Fig.~2]{clifton-93}.
More recently Hardy~\cite{Hardy-92,Hardy-93,hardy-97}
as well as Cabello and
Garc{\'{i}}a-Alcaine and
others~\cite{Cabello-1995-ppks,cabello-96,cabello-97-nhvp,Badziag-2011,Cabello-2013-HP,Cabello-2013-Hardylike} have discussed such scenarios.

Another true-implies-true configuration depicted in Fig.~\ref{2018-whycontexts-acsl1}(b)
has an immediate quantum realization~\cite[Table.~1, p.~102201-7]{2015-AnalyticKS} for
$\vert \langle {\bf a} \vert {\bf b}\rangle \vert^2=\frac{1}{2}$, and can be
extended to arbitrary  angles between non-collinear and non-orthogonal vectors.

A combo of Specker bug logics renders a non-separable set of two-valued states~\cite[$\Gamma_3$, p.~70]{kochen1}:
in the logic depicted in Fig.~\ref{2018-whycontexts-fbugcombo}(c),
${\bf a}$ and ${\bf a}'$
as well as
${\bf b}$ and ${\bf b}'$
cannot be ``separated'' from one another by any non-contextual (quasi)classical truth assignment enumerated in Appendix~\ref{2018-whycontexts-app-sbcombo}.
Kochen and Specker~\cite[Theorem~0, p.~67]{kochen1}  pointed out that
this type of inseparability is a necessary and sufficient condition
for a logic to be not embeddable in any classical Boolean algebra.
Therefore, whereas both the Specker bug logic as well as its extension true-implies-true logic can be represented by a partition logic,
the combo Specker bug logic cannot.

\subsection{Logics inducing partial value (in)definiteness}

Probably the strongest forms on value indefiniteness~\cite{pitowsky:218,hru-pit-2003}
are theorems~\cite{2012-incomput-proofsCJ,PhysRevA.89.032109,2015-AnalyticKS}
stating that relative to reasonable (admissibility, non-contextuality) assumptions,
if a quantized system is prepared in some pure state $\vert {\bf a} \rangle$,
then {\em any} observable which is not identical or orthogonal to $\vert {\bf a} \rangle$
is undefined. That is, there exist finite systems of quantum contexts
whose pasting are demanding that any pure state  $\vert {\bf b} \rangle$ not belonging to some context with  $\vert {\bf a} \rangle$
can {\em neither be true nor false}; else a complete contradiction would follow from the assumption of classically pre-existent truth values
on some pasting of contexts such as the Specker bug logic.

What does ``strong'' mean here?
Suppose one prepares the system in a particular context ${\cal C}$ such that a
single vector $\vert {\bf a} \rangle \in {\cal C}$ is true; that is, $\vert {\bf a} \rangle$ has probability measure $1$ when measured along ${\cal C}$.
Then, if one measures a complementary variable $\vert {\bf b} \rangle$,
and $\vert {\bf b} \rangle$ is sufficiently separated from $\vert {\bf a} \rangle$ (more precisely, at least an angle $\textrm{arccos} \left( \frac{1}{3} \right)$ apart
for the Specker bug logic),
then intertwined quantum propositional structures (such as the Specker bug logic) exist which,   interpreted (quasi)classically,
demand that  $\vert {\bf b} \rangle$ can never occur (cannot be true) -- and yet quantum system allow  $\vert {\bf b} \rangle$ to occur.
Likewise, other intertwined contexts which correspond to true-implies-true configurations of quantum observables
(termed Hardy-like~\cite{Hardy-92,Hardy-93,hardy-97}
by Cabello~\cite{Cabello-2013-Hardylike})
(quasi)classically imply
that some endpoint $\vert {\bf b}' \rangle$ must always occur, given  $\vert {\bf a} \rangle$ is true.
And yet, quantum mechanically, since $\vert {\bf a} \rangle$ and $\vert {\bf b}' \rangle$
are not collinear, quantum mechanics predicts that occasionally $\vert {\bf b}' \rangle$ does {\em not} occur.
In the ``strongest'' form~\cite{2012-incomput-proofsCJ,PhysRevA.89.032109,2015-AnalyticKS}  of classical ``do's and don'ts''
there are no possibilities whatsoever for an observable proposition to be either true or false.
That is, even if the Specker bug simultaneously allows some $\vert {\bf a} \rangle$  to be true and $\vert {\bf b} \rangle$ to be false
(although disallowing the latter to be true),
there is another,  supposedly more sophisticated finite configuration of intertwined quantum contexts, which can be constructively enumerated
and which disallows $\vert {\bf b} \rangle$
even to be false (it cannot be true either).

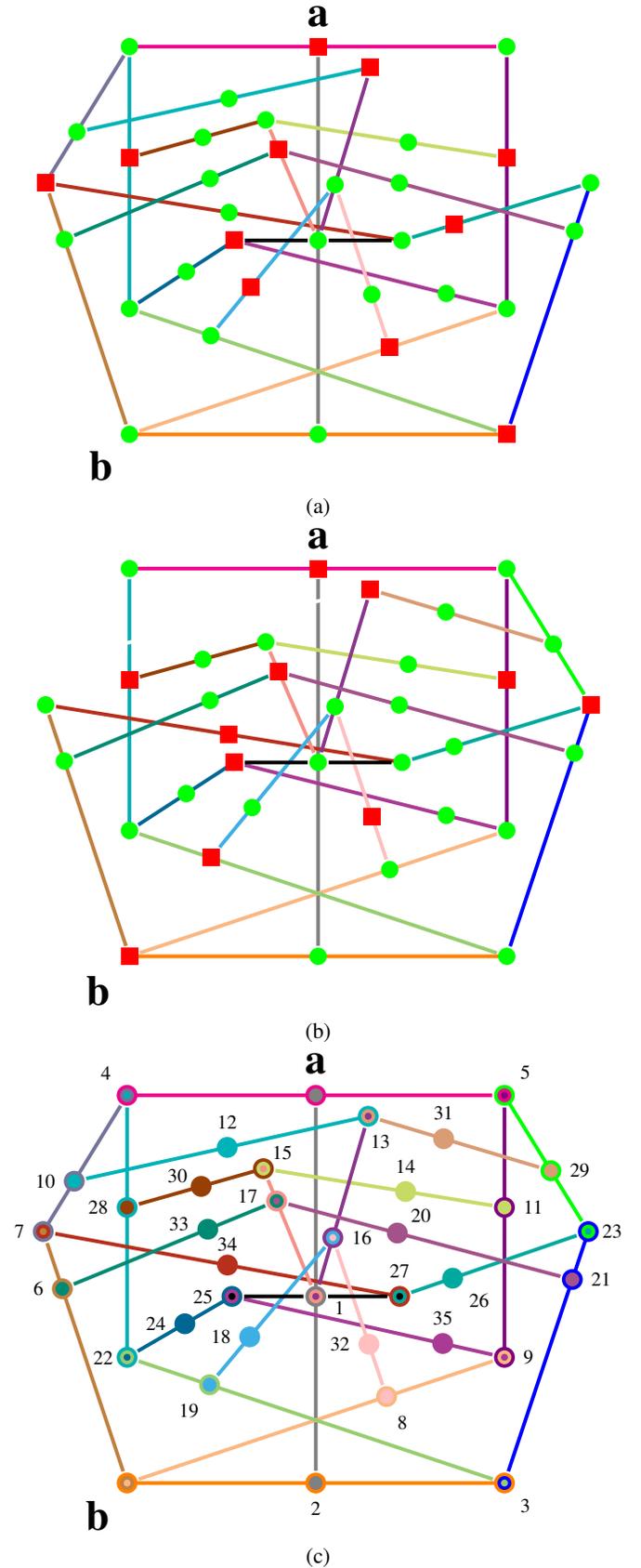
\begin{figure}
\begin{center}
\begin{tabular}{ccc}
\begin{tikzpicture} [scale=0.3]

   \tikzstyle{every path}=[line width=1.5pt]
   \tikzstyle{c1}=[color=green,circle,inner sep=2.5]
   \tikzstyle{s1}=[color=red,rectangle,inner sep=3.5]
   \tikzstyle{l1}=[draw=none,circle,minimum size=4]


\draw [color=orange] (4,0) coordinate[c1,fill,label=225:{\color{black}\LARGE ${\bf b}$}] (b) -- (13,0)  coordinate[c1,fill,label=270:{\iflabel \tiny $P_2$\fi}] (2) -- (22,0) coordinate[s1,fill,label=315:{\iflabel \tiny $P_3$\fi}] (3);
\draw [color=blue, ] (3) -- (26,12) coordinate[c1,fill,pos=0.8,label=0:{\iflabel \tiny $P_{21}$\fi}] (21) coordinate[c1,fill,label=0:{\iflabel \tiny $P_{23}$\fi}] (23);
\draw [color=white] (23) -- (22,18.5) coordinate[c1,fill,pos=0.4,color=white,label=0:{\iflabel \tiny $P_{29}$\fi}] (29) coordinate[c1,fill,label=45:{\iflabel \tiny $P_5$\fi}] (5);
\draw [color=magenta,] (5)-- (13,18.5)coordinate[s1,fill,label=90:{\color{black}\LARGE ${\bf a}$}] (a) -- (4,18.5) coordinate[c1,fill,label=135:{\iflabel \tiny $P_4$\fi}] (4);
\draw [color=CadetBlue, ] (4) -- (0,12) coordinate[c1,fill,pos=0.6,label=180:{\iflabel \tiny $P_{10}$\fi}] (10) coordinate[s1,fill,label=180:{\iflabel \tiny $P_7$\fi}] (7);
\draw [color=brown, ](7) -- (b)   coordinate[c1,fill,pos=0.2,label=180:{\iflabel \tiny $P_6$\fi}] (6);

   \draw [color=gray] (a) -- (2) coordinate[c1,fill,pos=0.5,label=315:{\iflabel \tiny $P_1$\fi}] (1);

   \draw [color=violet] (5) -- (22,6) coordinate[s1,fill,pos=0.4,label=0:{\iflabel \tiny $P_{11}$\fi}] (11) coordinate[c1,fill,label=0:{\iflabel \tiny $P_9$\fi}] (9);

\draw [color=Apricot] (9) -- (b) coordinate[s1,fill,pos=0.3,label=280:{\iflabel \tiny $P_8$\fi}] (8);

\draw [color=TealBlue] (4) -- (4,6) coordinate[s1,fill,pos=0.4,label=180:{\iflabel \tiny $P_{28}$\fi}] (28) coordinate[c1,fill,label=180:{\iflabel \tiny $P_{22}$\fi}] (22);
\draw [color=YellowGreen] (22) -- (3) coordinate[c1,fill,pos=0.2,label=260:{\iflabel \tiny $P_{19}$\fi}] (19);

   \coordinate (25) at ([xshift=-4cm]1);
   \coordinate (27) at ([xshift=4cm]1);

\draw [color=MidnightBlue] (22) -- (25) coordinate[c1,fill,pos=0.5,label=115:{\iflabel \tiny $P_{24}$\fi}] (24) coordinate[s1,fill,label=270:{\iflabel \tiny $P_{25}$\fi}] (25);
\draw [color=Mulberry] (25) -- (9) coordinate[c1,fill,pos=0.8,label=90:{\iflabel \tiny $P_{35}$\fi}] (35);

\draw [color=BrickRed] (7) -- (27) coordinate[c1,fill,pos=0.5,label=90:{\iflabel \tiny $P_{34}$\fi}] (34) coordinate[c1,fill,label=90:{\iflabel \tiny $P_{27}$\fi}] (27);
\draw [color=Emerald] (27) -- (23) coordinate[s1,fill,pos=0.25,label=270:{\iflabel \tiny $P_{26}$\fi}] (26);

\draw [color=BlueGreen] (10) -- (15.5,17.5) coordinate[c1,fill,pos=0.5,label=90:{\iflabel \tiny $P_{12}$\fi}] (12) coordinate[s1,fill,label=15:{\iflabel \tiny $P_{13}$\fi}] (13);

\draw [color=RawSienna] (28) -- (10.5,15) coordinate[c1,fill,pos=0.5,label=90:{\iflabel \tiny $P_{30}$\fi}] (30) coordinate[c1,fill,label=90:{\iflabel \tiny $P_{15}$\fi}] (15);
\draw [color=SpringGreen] (15) -- (11) coordinate[c1,fill,pos=0.6,label=90:{\iflabel \tiny $P_{14}$\fi}] (14);

\draw [color=Salmon] (15) -- (1) coordinate[s1,fill,pos=0.2,label=15:{\iflabel \tiny $P_{17}$\fi}] (17);
\draw [color=Fuchsia] (1)-- (13) coordinate[c1,fill,pos=0.3,label=0:{\iflabel \tiny $P_{16}$\fi}] (16);

\draw [color=CornflowerBlue] (19) -- (16) coordinate[s1,fill,pos=0.3,label=180:{\iflabel \tiny $P_{18}$\fi}] (18);
\draw [color=pink] (16) -- (8) coordinate[c1,fill,pos=0.7,label=180:{\iflabel \tiny $P_{32}$\fi}] (32);

\draw [color=PineGreen] (6) -- (17) coordinate[c1,fill,pos=0.7,label=90:{\iflabel \tiny $P_{33}$\fi}] (33);
\draw [color=DarkOrchid] (17) -- (21) coordinate[c1,fill,pos=0.4,label=90:{\iflabel \tiny $P_{20}$\fi}] (20);

\draw [color=black] (25) -- (1) -- (27);


\end{tikzpicture}
\\
(a)
\\
\begin{tikzpicture} [scale=0.3]

   \tikzstyle{every path}=[line width=1.5pt]
   \tikzstyle{c1}=[color=green,circle,inner sep=2.5]
   \tikzstyle{s1}=[color=red,rectangle,inner sep=3.5]
   \tikzstyle{l1}=[draw=none,circle,minimum size=4]


\draw [color=orange] (4,0) coordinate[s1,fill,label=225:{\color{black}\LARGE ${\bf b}$}] (b) -- (13,0)  coordinate[c1,fill,label=270:{\iflabel \tiny $P_2$\fi}] (2) -- (22,0) coordinate[c1,fill,label=315:{\iflabel \tiny $P_3$\fi}] (3);
\draw [color=blue, ] (3) -- (26,12) coordinate[c1,fill,pos=0.8,label=0:{\iflabel \tiny $P_{21}$\fi}] (21) coordinate[s1,fill,label=0:{\iflabel \tiny $P_{23}$\fi}] (23);
\draw [color=green] (23) -- (22,18.5) coordinate[c1,fill,pos=0.4,label=0:{\iflabel \tiny $P_{29}$\fi}] (29) coordinate[c1,fill,label=45:{\iflabel \tiny $P_5$\fi}] (5);
\draw [color=magenta,] (5)-- (13,18.5)coordinate[s1,fill,label=90:{\color{black}\LARGE ${\bf a}$}] (a) -- (4,18.5) coordinate[c1,fill,label=135:{\iflabel \tiny $P_4$\fi}] (4);
\draw [color=white] (4) -- (0,12) coordinate[c1,color=white,fill,pos=0.6,label=180:{\iflabel \tiny $P_{10}$\fi}] (10) coordinate[c1,fill,label=180:{\iflabel \tiny $P_7$\fi}] (7);
\draw [color=brown, ] (7) -- (b)   coordinate[c1,fill,pos=0.2,label=180:{\iflabel \tiny $P_6$\fi}] (6);

   \draw [color=gray] (a) -- (2) coordinate[c1,fill,pos=0.5,label=315:{\iflabel \tiny $P_1$\fi}] (1);

   \draw [color=violet] (5) -- (22,6) coordinate[s1,fill,pos=0.4,label=0:{\iflabel \tiny $P_{11}$\fi}] (11) coordinate[c1,fill,label=0:{\iflabel \tiny $P_9$\fi}] (9);

\draw [color=Apricot] (9) -- (b) coordinate[c1,fill,pos=0.3,label=280:{\iflabel \tiny $P_8$\fi}] (8);

\draw [color=TealBlue] (4) -- (4,6) coordinate[s1,fill,pos=0.4,label=180:{\iflabel \tiny $P_{28}$\fi}] (28) coordinate[c1,fill,label=180:{\iflabel \tiny $P_{22}$\fi}] (22);
\draw [color=YellowGreen] (22) -- (3) coordinate[s1,fill,pos=0.2,label=260:{\iflabel \tiny $P_{19}$\fi}] (19);

   \coordinate (25) at ([xshift=-4cm]1);
   \coordinate (27) at ([xshift=4cm]1);

\draw [color=MidnightBlue] (22) -- (25) coordinate[c1,fill,pos=0.5,label=115:{\iflabel \tiny $P_{24}$\fi}] (24) coordinate[s1,fill,label=270:{\iflabel \tiny $P_{25}$\fi}] (25);
\draw [color=Mulberry] (25) -- (9) coordinate[c1,fill,pos=0.8,label=90:{\iflabel \tiny $P_{35}$\fi}] (35);

\draw [color=BrickRed] (7) -- (27) coordinate[s1,fill,pos=0.5,label=90:{\iflabel \tiny $P_{34}$\fi}] (34) coordinate[c1,fill,label=90:{\iflabel \tiny $P_{27}$\fi}] (27);
\draw [color=Emerald] (27) -- (23) coordinate[c1,fill,pos=0.25,label=270:{\iflabel \tiny $P_{26}$\fi}] (26);

\draw [color=white] (10) -- (15.5,17.5) coordinate[c1,color=white,fill,pos=0.5,label=90:{\iflabel \tiny $P_{12}$\fi}] (12) coordinate[s1,fill,label=15:{\iflabel \tiny $P_{13}$\fi}] (13);
\draw [color=Tan] (13) -- (29) coordinate[c1,fill,pos=0.4,label=90:{\iflabel \tiny $P_{31}$\fi}] (31);

\draw [color=RawSienna] (28) -- (10.5,15) coordinate[c1,fill,pos=0.5,label=90:{\iflabel \tiny $P_{30}$\fi}] (30) coordinate[c1,fill,label=90:{\iflabel \tiny $P_{15}$\fi}] (15);
\draw [color=SpringGreen] (15) -- (11) coordinate[c1,fill,pos=0.6,label=90:{\iflabel \tiny $P_{14}$\fi}] (14);

\draw [color=Salmon] (15) -- (1) coordinate[s1,fill,pos=0.2,label=15:{\iflabel \tiny $P_{17}$\fi}] (17);
\draw [color=Fuchsia] (1)-- (13) coordinate[c1,fill,pos=0.3,label=0:{\iflabel \tiny $P_{16}$\fi}] (16);

\draw [color=CornflowerBlue] (19) -- (16) coordinate[c1,fill,pos=0.3,label=180:{\iflabel \tiny $P_{18}$\fi}] (18);
\draw [color=pink] (16) -- (8) coordinate[s1,fill,pos=0.7,label=180:{\iflabel \tiny $P_{32}$\fi}] (32);

\draw [color=PineGreen] (6) -- (17) coordinate[c1,fill,pos=0.7,label=90:{\iflabel \tiny $P_{33}$\fi}] (33);
\draw [color=DarkOrchid] (17) -- (21) coordinate[c1,fill,pos=0.4,label=90:{\iflabel \tiny $P_{20}$\fi}] (20);

\draw [color=black] (25) -- (1) -- (27);

   \coordinate (ContextLabel) at ([shift=({-2cm,-3mm})]1);
   \draw (ContextLabel) coordinate[l1,label=90:{\iflabel \tiny $C_{26}$\fi}];

   \end{tikzpicture}
\\
(b)
\\
\begin{tikzpicture}  [scale=0.3]
\labeltrue
        \tikzstyle{every path}=[line width=1.5pt]
        \tikzstyle{c1}=[circle,fill,inner sep=3]
        \tikzstyle{c2}=[circle,fill,inner sep=2]
        \tikzstyle{c3}=[circle,fill,inner sep=1]
        \tikzstyle{s1}=[color=red,rectangle,minimum size=8,inner sep=6]
        \tikzstyle{d1}=[draw=none,circle,minimum size=4]
        \tikzstyle{e1}=[color=gray,rectangle,minimum size=8,inner sep=6]


\draw [color=orange]  (4,0)  coordinate[c1,fill,label=225:{\color{black}\LARGE ${\bf b}$}] (b) -- (13,0)    coordinate[c1,fill,label={[label distance=-1]270:{\iflabel \footnotesize \color{black}  $2$\fi}}] (2) -- (22,0)  coordinate[c1,fill,label={[label distance=-1]315:{\iflabel \footnotesize \color{black}  $3$\fi}}] (3);
\draw [color=blue] (3) -- (26,12)  coordinate[c1,fill,pos=0.8,label={[label distance=-1]0:{\iflabel \footnotesize \color{black}  ${21}$\fi}}] (21) coordinate[c1,fill,label={[label distance=-3]0:{\iflabel \footnotesize \color{black}  ${23}$\fi}}] (23);
\draw [color=green] (23) -- (22,18.5) coordinate[c1,fill,pos=0.4,label={[label distance=-1]0:{\iflabel \footnotesize \color{black}  ${29}$\fi}}] (29) coordinate[c1,fill,label={[label distance=-1]45:{\iflabel \footnotesize \color{black}  $5$\fi}}] (5);
\draw [color=magenta] (5)-- (13,18.5)coordinate[c1,fill,label=90:{\color{black}\LARGE ${\bf a}$}] (a) -- (4,18.5)  coordinate[c1,fill,label={[label distance=-1]135:{\iflabel \footnotesize \color{black}  $4$\fi}}] (4);
\draw [color=CadetBlue] (4) -- (0,12)   coordinate[c1,fill,pos=0.6,label={[label distance=-1]180:{\iflabel \footnotesize \color{black}  ${10}$\fi}}] (10)  coordinate[c1,fill,label={[label distance=-1]180:{\iflabel \footnotesize \color{black}  $7$\fi}}] (7);
\draw [color=brown] (7) -- (b)      coordinate[c1,fill,pos=0.2,label={[label distance=-1]180:{\iflabel \footnotesize \color{black}  $6$\fi}}] (6);

        \draw [color=gray] (a) -- (2) coordinate[c1,fill,pos=0.52,label={[label distance=-1, yshift=2]357.5:{\iflabel \footnotesize \color{black}  $1$\fi}}] (1);

        \draw [color=violet] (5) -- (22,6) coordinate[c1,fill,pos=0.4,label={[label distance=-1]0:{\iflabel \footnotesize \color{black}  ${11}$\fi}}] (11) coordinate[c1,fill,label={[label distance=-1]0:{\iflabel \footnotesize \color{black}  $9$\fi}}] (9);

\draw [color=Apricot] (9) -- (b) coordinate[c1,fill,pos=0.3,label={[label distance=-1]280:{\iflabel \footnotesize \color{black}  $8$\fi}}] (8);

\draw [color=TealBlue] (4) -- (4,6) coordinate[c1,fill,pos=0.4,label={[label distance=-1]180:{\iflabel \footnotesize \color{black}  ${28}$\fi}}] (28) coordinate[c1,fill,label={[label distance=-3]180:{\iflabel \footnotesize \color{black}  ${22}$\fi}}] (22);
\draw [color=YellowGreen] (22) -- (3) coordinate[c1,fill,pos=0.2,label={[label distance=-1]260:{\iflabel \footnotesize \color{black}  ${19}$\fi}}] (19);

        \coordinate (25) at ([xshift=-4cm]1);
        \coordinate (27) at ([xshift=4cm]1);

\draw [color=MidnightBlue]  (22) -- (25) coordinate[c1,fill,pos=0.5,label={[label distance=-1]180:{\iflabel \footnotesize \color{black}  ${24}$\fi}}] (24) coordinate[c1,fill,label={[label distance=-1]180:{\iflabel \footnotesize \color{black}  ${25}$\fi}}] (25);
\draw [color=Mulberry] (25) -- (9) coordinate[c1,fill,pos=0.8,label={[label distance=-1]90:{\iflabel \footnotesize \color{black}  ${35}$\fi}}] (35);

\draw [color=BrickRed]  (7) -- (27) coordinate[c1,fill,pos=0.5,label={[label distance=-3]90:{\iflabel \footnotesize \color{black}  ${34}$\fi}}] (34) coordinate[c1,fill,label={[label distance=-1]90:{\iflabel \footnotesize \color{black}  ${27}$\fi}}] (27);
\draw [color=Emerald] (27) -- (23) coordinate[c1,fill,pos=0.25,label={[label distance=-1]320:{\iflabel \footnotesize \color{black}  ${26}$\fi}}] (26);

\draw [color=BlueGreen]  (10) -- (15.5,17.5) coordinate[c1,fill,pos=0.5,label={[label distance=-1]90:{\iflabel \footnotesize \color{black}  ${12}$\fi}}] (12) coordinate[c1,fill,label={[label distance=-1,xshift=5]270:{\iflabel \footnotesize \color{black}  ${13}$\fi}}] (13);
\draw [color=Tan] (13) -- (29) coordinate[c1,fill,pos=0.4,label={[label distance=-1]90:{\iflabel \footnotesize \color{black}  ${31}$\fi}}] (31);

\draw [color=RawSienna]  (28) -- (10.5,15) coordinate[c1,fill,pos=0.5,label={[label distance=-3, yshift=-3]160:{\iflabel \footnotesize \color{black}  ${30}$\fi}}] (30) coordinate[c1,fill,label={[label distance=-5]45:{\iflabel \footnotesize \color{black}  ${15}$\fi}}] (15);
\draw [color=SpringGreen] (15) -- (11) coordinate[c1,fill,pos=0.6,label={[label distance=-1]90:{\iflabel \footnotesize \color{black}  ${14}$\fi}}] (14);

\draw [color=Salmon]  (15) -- (1) coordinate[c1,fill,pos=0.2,label={[label distance=-1, yshift=2]180:{\iflabel \footnotesize \color{black}  ${17}$\fi}}] (17);
\draw [color=Fuchsia] (1)-- (13) coordinate[c1,fill,pos=0.3,label={[label distance=-1]0:{\iflabel \footnotesize \color{black}  ${16}$\fi}}] (16);

\draw [color=CornflowerBlue]  (19) -- (16) coordinate[c1,fill,pos=0.3,label={[label distance=-1]180:{\iflabel \footnotesize \color{black}  ${18}$\fi}}] (18);
\draw [color=pink] (16) -- (8) coordinate[c1,fill,pos=0.7,label={[label distance=-1]180:{\iflabel \footnotesize \color{black}  ${32}$\fi}}] (32);

\draw [color=PineGreen]  (6) -- (17) coordinate[c1,fill,pos=0.7,label={[label distance=-1, yshift=2]180:{\iflabel \footnotesize \color{black}  ${33}$\fi}}] (33);
\draw [color=DarkOrchid] (17) -- (21) coordinate[c1,fill,pos=0.4,label={[label distance=-3]20:{\iflabel \footnotesize \color{black}  ${20}$\fi}}] (20);

\draw [color=black] (25) -- (1) -- (27);

\draw (a) coordinate[c2,fill=gray];

\draw (b) coordinate[c2,fill=brown];
\draw (b) coordinate[c3,fill=Apricot];

\draw (4) coordinate[c2,fill=CadetBlue];
\draw (4) coordinate[c3,fill=TealBlue];

\draw (5) coordinate[c2,fill=magenta];
\draw (5) coordinate[c3,fill=violet];

\draw (7) coordinate[c2,BrickRed];
\draw (7) coordinate[c3,fill=brown];

\draw (23) coordinate[c2,fill=green];
\draw (23) coordinate[c3,fill=Emerald];

\draw (29) coordinate[c2,fill=Tan];

\draw (21) coordinate[c2,fill=DarkOrchid];

\draw (3) coordinate[c2,fill=blue];
\draw (3) coordinate[c3,fill=YellowGreen];

\draw (11) coordinate[c2,fill=SpringGreen];

\draw (9) coordinate[c2,fill=Apricot];
\draw (9) coordinate[c3,fill=Mulberry];

\draw (27) coordinate[c2,fill=Emerald];
\draw (27) coordinate[c3,fill=black];

\draw (13) coordinate[c2,fill=Tan];
\draw (13) coordinate[c3,fill=Fuchsia];

\draw (7) coordinate[c2,BrickRed];
\draw (7) coordinate[c3,fill=brown];

\draw (8) coordinate[c2,fill=pink];

\draw (19) coordinate[c2,fill=CornflowerBlue];

\draw (16) coordinate[c2,fill=CornflowerBlue];
\draw (16) coordinate[c3,fill=pink];

\draw (1) coordinate[c2,fill=Salmon];
\draw (1) coordinate[c3,fill=Fuchsia];

\draw (17) coordinate[c2,fill=PineGreen];
\draw (17) coordinate[c3,fill=DarkOrchid];

\draw (15) coordinate[c2,fill=SpringGreen];
\draw (15) coordinate[c3,fill=Salmon];

\draw (25) coordinate[c2,fill=Mulberry];
\draw (25) coordinate[c3,fill=black];

\draw (22) coordinate[c2,fill=YellowGreen];
\draw (22) coordinate[c3,fill=MidnightBlue];

\draw (28) coordinate[c2,fill=RawSienna];

\draw (10) coordinate[c2,fill=BlueGreen];

\draw (6) coordinate[c2,fill=PineGreen];

\draw (2) coordinate[c2,fill=gray];

\end{tikzpicture}
\\
(c)
\end{tabular}
\end{center}
\caption{Greechie orthogonality diagram of a logic~\cite[Fig.~2, p.~102201-8]{2015-AnalyticKS}
realizable in $\mathbb{R}^3$
(a) with the true--implies--false property;
(b) with the true--implies--true property;
(c) with the true--implies--value indefiniteness (neither true nor false) property on the atoms ${\bf a}$ and ${\bf b}$,
respectively. (a) and (b) contain the single (out of 13) value assignment which is possible, and for which ${\bf a}$ is true.
All 8 value assignments of the logic depicted in (c) require ${\bf a}$ to be false.
}
\label{2018-whycontexts-acsl1}
\end{figure}

For the sake of an explicit example take the logic~\cite[Fig.~2, p.~102201-8]{2015-AnalyticKS}
depicted in Fig.~\ref{2018-whycontexts-acsl1}(c).
It is the composite of two logics depicted in Figs.~\ref{2018-whycontexts-acsl1}(a),(b),
which perform very differently at ${\bf b}$ given ${\bf a}$ to be true:
whereas (a) implements a true-implies-false property,
(b) has a true-implies-true property for the atoms ${\bf a}$ and ${\bf b}$, respectively.
Both (a) and (b) are proper subsets (lacking 2 contexts)
of the logic in Fig.~\ref{2018-whycontexts-acsl1}(c);
and, apart from their difference in 4 contexts, are identical.

More precisely, as explicated
in Appendix~\ref{2018-whycontexts-app-ACS},
both of these logics (a) and (b) allow 13 truth assignments (two-valued states), but only a single one allows ${\bf a}$ to be true on either of them.
(This uniqueness is not essential to the argument.)
The logic in (c) allows for 8 truth assignments, but all of them assign falsity to ${\bf a}$.
By combining the logics (a) and (b) one obtains (c) which, if ${\bf a}$ is assumed to be true, implies that ${\bf b}$
can neither be true
-- this would contradict the true-implies-false property of (a) --
nor can it be false
-- because this would contradict the true-implies-true property of (b).
Hence we are left with the only consistent alternative, (relative to the assumptions):
that a system prepared in state ${\bf a}$ must be  value indefinite for observable ${\bf b}$.
Thereby, as the truth assignment on ${\bf b}$ is not defined, it must be partial on the entire logic depicted in Fig.~\ref{2018-whycontexts-acsl1}(c).

\newif\iflabel \labelfalse

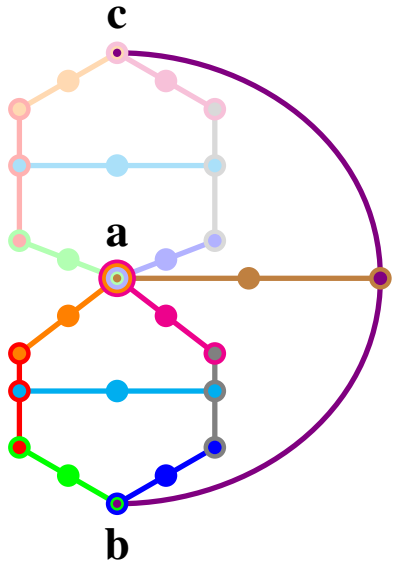
\begin{figure}
\begin{center}
\begin{tikzpicture}  [xscale=0.25, yscale=0.25, rotate=0]

\newdimen\ms
\ms=0.1cm

\tikzstyle{every path}=[line width=2pt]

\tikzstyle{c5}=[circle,inner sep={\ms/8},minimum size=5*\ms]
\tikzstyle{c4}=[circle,inner sep={\ms/8},minimum size=4*\ms]
\tikzstyle{c3}=[circle,inner sep={\ms/8},minimum size=3*\ms]
\tikzstyle{c2}=[circle,inner sep={\ms/8},minimum size=1.8*\ms]
\tikzstyle{c1}=[circle,inner sep={\ms/8},minimum size=1.1*\ms]

\newdimen\R
\R=6cm     


\newdimen\K
\K=12cm
\newdimen\KK
\KK=11cm
\newdimen\KKK
\KKK=11.5cm

\newdimen\RR
\RR=-1cm
\newdimen\RRR
\RRR=-0.5cm

\path
  ({90 + 0 * 360 /6}:\R      )   coordinate(1)
  ({90 + 30 + 0 * 360 /6}:{\R * sqrt(3)/2}      ) + (0,\RRR)   coordinate(2)
  ({90 + 1 * 360 /6}:\R   )  + (0,\RR)  coordinate(3)
  ({90 + 30 + 1 * 360 /6}:{\R * sqrt(3)/2}   ) coordinate(4)
  ({90 + 2 * 360 /6}:\R  ) coordinate(5)
  ({90 + 30 + 2 * 360 /6}:{\R * sqrt(3)/2}  ) coordinate(6)
  ({90 + 3 * 360 /6}:\R  ) coordinate(7)
  ({90 + 30 + 3 * 360 /6}:{\R * sqrt(3)/2}  ) coordinate(8)
  ({90 + 4 * 360 /6}:\R     ) coordinate(9)
  ({90 + 30 + 4 * 360 /6}:{\R * sqrt(3)/2}     ) coordinate(10)
  ({90 + 5 * 360 /6}:\R     ) + (0,\RR)  coordinate(11)
  ({90 + 30 + 5 * 360 /6}:{\R * sqrt(3)/2}     ) + (0,\RRR)   coordinate(12)
  (0,0) coordinate(13)

  ({90 + 0 * 360 /6}:\R     ) + (0,\K)  coordinate(21)
  ({90 + 30 + 0 * 360 /6}:{\R * sqrt(3)/2}     ) + (0,\K)  coordinate(22)
  ({90 + 1 * 360 /6}:\R  ) + (0,\K)  coordinate(23)
  ({90 + 30 + 1 * 360 /6}:{\R * sqrt(3)/2}  ) + (0,\K)  coordinate(24)
  ({90 + 2 * 360 /6}:\R  ) + (0,\KK)  coordinate(25)
  ({90 + 30 + 2 * 360 /6}:{\R * sqrt(3)/2}  ) + (0,\KKK)  coordinate(26)
  ({90 + 3 * 360 /6}:\R  ) + (0,\K)  coordinate(27)
  ({90 + 30 + 3 * 360 /6}:{\R * sqrt(3)/2}  ) + (0,\KKK)  coordinate(28)
  ({90 + 4 * 360 /6}:\R    ) + (0,\KK)  coordinate(29)
  ({90 + 30 + 4 * 360 /6}:{\R * sqrt(3)/2}    ) + (0,\K)  coordinate(210)
  ({90 + 5 * 360 /6}:\R    ) + (0,\K)  coordinate(211)
  ({90 + 30 + 5 * 360 /6}:{\R * sqrt(3)/2}    ) + (0,\K)  coordinate(212)
  (0,0) + (0,\K) coordinate(213)

   (0,0) + (14,{\K/2}) coordinate(40)
   (0,0) + (7,{\K/2}) coordinate(41)
;


\draw [color=orange] (1) -- (2) -- (3);
\draw [color=red] (3) -- (4) -- (5);
\draw [color=green] (5) -- (6) -- (7);
\draw [color=blue] (7) -- (8) -- (9);
\draw [color=gray] (9) -- (10) -- (11);    %
\draw [color=magenta] (11) -- (12) -- (1);    %
\draw [color=cyan] (4) -- (10);

\draw [color=orange!30!white] (21) -- (22) -- (23);
\draw [color=red!30!white] (23) -- (24) -- (25);
\draw [color=green!30!white] (25) -- (26) -- (27);
\draw [color=blue!30!white] (27) -- (28) -- (29);
\draw [color=gray!30!white] (29) -- (210) -- (211);    %
\draw [color=magenta!30!white] (211) -- (212) -- (21);    %
\draw [color=cyan!30!white] (24) -- (210);

\draw [color=violet] (7)  to   [out=0,in=270] (40) to [out=90,in=0] (21);
\draw [color=brown] (27)  -- (40);

%
%

\draw (1) coordinate[c5,fill=magenta,label=90:\LARGE ${\bf a}$];   %
\draw (1) coordinate[c4,fill=orange];  %
\draw (1) coordinate[c1,fill=brown];  %
\draw (2) coordinate[c3,fill=orange];    %
\draw (3) coordinate[c3,fill=red]; %
\draw (3) coordinate[c2,fill=orange];  %
\draw (4) coordinate[c3,fill=red];  %
\draw (4) coordinate[c2,fill=cyan];  %
\draw (5) coordinate[c3,fill=green];  %
\draw (5) coordinate[c2,fill=red];  %
\draw (6) coordinate[c3,fill=green];
\draw (7) coordinate[c3,fill=blue,label=270:\LARGE ${\bf b}$];  %
\draw (7) coordinate[c2,fill=green];  %
\draw (7) coordinate[c1,fill=violet];  %
\draw (8) coordinate[c3,fill=blue];  %
\draw (9) coordinate[c3,fill=gray];
\draw (9) coordinate[c2,fill=blue];  %
\draw (10) coordinate[c3,fill=gray];  %
\draw (10) coordinate[c2,fill=cyan];  %
\draw (11) coordinate[c3,fill=magenta];  %
\draw (11) coordinate[c2,fill=gray]; %
\draw (12) coordinate[c3,fill=magenta];  %
\draw (13) coordinate[c3,fill=cyan];  %

\draw (21) coordinate[c3,fill=magenta!30!white,label=90:\LARGE ${\bf c}$];   
\draw (21) coordinate[c2,fill=orange!30!white];  %
\draw (21) coordinate[c1,fill=violet];  %
\draw (22) coordinate[c3,fill=orange!30!white];    %
\draw (23) coordinate[c3,fill=red!30!white]; %
\draw (23) coordinate[c2,fill=orange!30!white];  %
\draw (24) coordinate[c3,fill=red!30!white];  %
\draw (24) coordinate[c2,fill=cyan!30!white];  %
\draw (25) coordinate[c3,fill=green!30!white];  %
\draw (25) coordinate[c2,fill=red!30!white];  %
\draw (26) coordinate[c3,fill=green!30!white];
\draw (27) coordinate[c3,fill=blue!30!white,label=270:{}];  %
\draw (27) coordinate[c2,fill=green!30!white];  %
\draw (27) coordinate[c1,fill=brown];  %
\draw (28) coordinate[c3,fill=blue!30!white];  %
\draw (29) coordinate[c3,fill=gray!30!white];
\draw (29) coordinate[c2,fill=blue!30!white];  %
\draw (210) coordinate[c3,fill=gray!30!white];  %
\draw (210) coordinate[c2,fill=cyan!30!white];  %
\draw (211) coordinate[c3,fill=magenta!30!white];  %
\draw (211) coordinate[c2,fill=gray!30!white]; %
\draw (212) coordinate[c3,fill=magenta!30!white];  %
\draw (213) coordinate[c3,fill=cyan!30!white];  %

\draw (40) coordinate[c3,fill=brown,label=0:\LARGE $$];  
\draw (40) coordinate[c2,fill=violet];  %

\draw (41) coordinate[c3,fill=brown,label=290:\LARGE $$];  
\end{tikzpicture}
\end{center}
\caption{Greechie orthogonality diagram of a logic which is value indefinite on
${\bf b}$ (as well as on ${\bf c}$ for symmetry reasons), given ${\bf a}$ is true;
alas such a logic has no realization in three dimensional Hilbert space, as the
angles
$\angle  ({\bf a},{\bf b})$
between ${\bf a}$ and ${\bf b}$
should simultaneously obey
$  1.2 \approx \textrm{arcsec} (3) \le  \angle  ({\bf a},{\bf b}) \le   \textrm{arccsc} (3)\approx 0.3$.}
\label{2018-whycontexts-impossible}
\end{figure}

The scheme of the proof is as follows:
\begin{itemize}
\item[(i)]
Find a logic (collection of intertwined contexts of observables) exhibiting a true-implies-false  property on the  two atoms ${\bf a}$ and ${\bf b}$.
\item[(ii)]
Find another logic exhibiting a true-implies-true property on the same two atoms ${\bf a}$ and ${\bf b}$.
\item[(iii)]
Then join (paste) these logics into a larger logic, which, given ${\bf a}$,
neither allows ${\bf b}$ to be true nor false. Consequently  ${\bf b}$ must be value indefinite.
\end{itemize}
The most suggestive candidate for such a pasting
is, however, unavailable:
it is the combination of a Specker bug logic  and another, extended Specker bug logic, as depicted in Fig.~\ref{2018-whycontexts-impossible}.
Such logic cannot be realized in three dimensions, as the angles cannot be chosen consistently; that is, obeying the Cabello bounds on the relative angles, respectively.

The latter result about the partiality of the truth assignment
has already been discussed by Pitowsky~\cite{pitowsky:218}, and later by Hrushovski and Pitowsky~\cite{hru-pit-2003}.
It should also be mentioned that the logic (c) has been realized with a particular configuration
in three-dimensional real Hilbert space~\cite[Tables~I,II, p.~102201-7]{2015-AnalyticKS} which are an angle $\angle ({\bf a} ,  {\bf b})=\textrm{arccos}\left(\frac{1}{\sqrt{2}}\right)$
apart, but,
as has been mentioned earlier,
this kind of value indefiniteness on any particular state ${\bf b}$, given that the system has been prepared in state ${\bf a}$, can be constructively
obtained by an extension of the above configuration whenever ${\bf a}$ and ${\bf b}$ are neither collinear (in this case ${\bf b}$ would be true)
nor orthogonal (in this case ${\bf b}$ would be false).
So basically all states not identical (or orthogonal) to the state prepared must be value indefinite.

All three logics in Fig.~\ref{2018-whycontexts-impossible}(a)-(c) have another non-classical feature: they are non-unital~\cite{tkadlec-96},
meaning that the truth assignments on some of their atoms can only acquire the value false, regardless of the preparation.
That is, in this ``state-independend'' form, whenever a proposition corresponding to such an atom is measured to be true,
this can be interpreted as indication of non-classicality.
(Note that one can always rotate the entire set of rays so that this particular atom coincides with some observable measured.)

\section{Propositional logic does not uniquely determine probabilities}
\label{2018-whycontexts-sec5}

By now it should be clear that the propositional structure does in general {\em not} uniquely determine its probabilities.
The Specker bug in Fig.~\ref{2018-whycontexts-fbugcombo}(a) serves as a good example for that:
it supports (quasi)classical probabilities,
explicitly enumerated in Refs.~\cite[p.~286, Fig.~11.5(iii)]{svozil-2016-s} and~\cite[p.~91, Fig.~12.10]{svozil-2016-pu-book}
which are formed by convex combinations of all two-valued states on them.

Other propositional structures such as the pentagon logic support ``exotic'' probability measures~\cite{wright:pent}
which do not vanish at their interlink observables and are equally weighted with value $\frac{1}{2}$ there.
This measure is neither realized in the (quasi)classical partition logic setup explicitly discussed in
Refs.~\cite[p.~289, Fig.~11.8]{svozil-2016-s} and~\cite[p.~88, Fig.~12.8]{svozil-2016-pu-book},
nor in quantum mechanics.
It remains to be seen if a more general theory of probability measures based on the axioms
{\bf A1}-{\bf A3} can be found.

\section{Some Platonist afterthoughts}
\label{2018-whycontexts-summary}

The author's not-so-humble reading of all these aforementioned ``mind-boggling'' non-classical quantum predictions is a rather sober one:
in view of the numerous indications that classical value definiteness
cannot be extended to more than a single context,  the most plausible supposition is that,
besides exotic possibilities~\cite{pitowsky-83,meyer:99}, ontologically there is only one such {\it ``Realding''}
-- indeed a rather obvious candidate suggesting itself as ontology:
a single vector, or rather a single context.
Quantized systems can be completely and exhaustively characterized by a unique context, and a ``true'' proposition within this context.

Suppose for a moment that this hypothesis is correct, and that there is no ontology, no {\it ``Realding,''} beyond a single context.
There is one preferred view -- namely the context identical to the context in which the system has been prepared -- and all but one epistemic views.

And yet,
a confusing experience is the apparent ease by which an experimenter appears to measure, without any difficulty, a context or (maximal) observable not (or only partly through intertwines) matching the preparation context.
In such a situation one may assume that the measurement grants an ``imperfect'' view on the preparation context.
In this process, information -- in particular, the relative locatedness of the measurement context with respect to the preparation context --
is {\em augmented} by properties of the measurement device, thereby effectively generating entanglement~\cite{schrodinger,london-Bauer-1983}
{\it via} context translation~\cite{svozil-2003-garda}.
Frames of reference which do not coincide with the {\it ``Realding''} or preparation context
necessarily include stochastic elements which are not caused or determined by any property of the formerly individual {\it ``Realding.''}
One may conclude~\cite{Howard2004-HOWWIT} with Bohr's 1972 Como lecture~\cite[p.~580]{bohr:28} that
{\em ``any observation of atomic
phenomena will involve an interaction with the
agency of observation not to be neglected.
Accordingly, an independent reality in the ordinary physical sense can neither be ascribed to the phenomena nor
to the agencies of observation.''}
That is, any interaction between the previously
separated individual object and the measurement device
results in a joint physical state which is
no longer determined by the states of the (previously) individual constituents~\cite{v-neumann-49,london-Bauer-1983}.
Instead, the joint state exhibits what Schr\"odinger later called  entanglement~\cite{schrodinger}.
Entanglement characterized by a value definite {\em relational}~\cite{zeil-99} or {\em collective} (re-)encoding of information with respect to the constituent parts,
thereby (since the unitary quantum evolution is injective) resulting in a value indefiniteness of the previously individual and separate parts.
As a result, knowledge about observations obtained by different contexts than the preparation context are necessarily
(at least partially in the sense of the augmented information from the measurement device) epistemic.

Another possible source of perplexity might be the various {\em types} of algabaic or logical structures involved.
Classically, empirical logics are Boolean algebras.
Then, in a first step towards non-classicality, there are partition logics which are not Boolean any longer
(they feature complementarity through non-distributivity)
but nevertheless still allow for a certain type of (quasi)classicality; that is, a separating and unital set of two-valued states.
And then, further on this road, there are (finite) quantum logics which do not allow any definite state at all.

One might be puzzled by the fact that there exist ``intermediate'' logics,
such as the Specker bug or the pentagon (pentagram) logic discussed in Sects.~\ref{2018-whycontexts-Specker-bug}
and~\ref{2018-whycontexts-pentagons},
which still allow (even classical) simultaneous value indefiniteness, although they contain
observables which are mutually complementary (non-collinear and non-orthogonal).
However, this apparent paradox should rather be interpreted epistemically, as means (configuration) relative~\cite{Myrvold2011237}:
in the case of the pentagon we have decided to concentrate on 10 observables in a cyclic pasting of 5 contexts,
but we have thereby implicitly chosen to ``look the other way''
and disregard the abundance of other observables which impose much more stringent conditions
on the value definiteness of the observables in the pentagon logic than the pentagon logic itself.

So, properties such as the true-implies-false, the true-implies-true properties, as well as inseparability and even value indefiniteness,
are means relative and valid only if one restricts or broadens one's attention to sometimes very specific, limited sublogics
of the realm of all conceivable quantum logics, which are structures
formed by perpendicular projection operators in Hilbert spaces of dimension larger than two.

Pointedly stated, sets of intertwining contexts connecting two (or more) relevant complementary observables
${\bf a}$ and
${\bf b}$
should be considered as totally arbitrary when it comes to the inclusion or exclusion of particular contexts interconnecting them:
there is neither a necessity nor even a compelling reason
to take into account one such structue and disregard another, or favour one over the other.
Indeed, in an extreme, {\it sui generis}, form of the argument,
suppose a single quantum is prepared in some state ${\bf a}$.
Then
{\em every single} outcome of a measurement of {\em every} complementary (non-collinear and non-orthogonal relative to the state prepared) quantum observable
may be considered as ``proof'' or ``certification of non-classicality'' (or, in another terminology, ``contextuality'').
Because  those observable can be identified with the ``endpoint'' ${\bf b}$ of either some true-implies-false,
or alternatively true-implies-true
configuration [say the one sketched in Figs.~\ref{2018-whycontexts-impossible}(a),(b)],
depending on whether the classical false or true predictions need to contradict the particular outcome, respectively.
For quantum logics with a unital set of two-valued states -- such as the logics depicted by Tkadlec~\cite[p.~207, Fig.~2]{tkadlec-96}
or the ones in Figs.~\ref{2018-whycontexts-impossible}(a),(b) --
one could even get rid of the state preparation if ${\bf b}$ occurs and is identified with an observable which, according to the classical predictions
associated with that logic, cannot occur.
There is no principle which could prevent us from arguing that way if we insist on the simultaneous existence of multiple contexts
encountered in quantum mechanics.
Indeed, are intertwining contexts not scholastic~\cite{specker-60} sophisms in desperate need of deconstruction?

An interesting historical question arises:
Kochen and Specker, in a succession of papers on partial algebras~\cite{kochen2,kochen3,kochen1}
have insisted that logical operations should only be defined within contexts and must not be applied to propositions outside of it.
And yet  they have considered extended counterfactual structures of pasted context,
ending up in a holistic argument involving complementary observables.
Of course, an immediate reply might be that without intertwined contexts there cannot be any
non-trivial (non-classical, non-Boolean) configuration of observables which is of any interest.

For the reasons mentioned earlier,
the emphasis should not be on ``completing'' quantum mechanics by some sort of hidden parameter theory
-- such as, for instance, Valentini~\cite{Valentini-2018} envisioning a theory which is to quantum mechanics
as statistical physics is to thermodynamics --
but just the opposite:
the challenge is to acknowledge the scarcity of resources --
the {\it ``Realding''} or physical state as mere vector --
despite the continuum of possible views on it,
resulting in an illusory over-abundance and over-determination.

In this line of thought
the question of what might be the reason behind the futility to co-define non-commuting quantum observables
(from two or more different contexts)
simultaneously should be answered
in terms of a serious lack of a proper perspective of what one is dealing with:
Metaphorically speaking, it is almost as if one pretends to take a $360^\circ$ panorama of what lies in the outside world
while actually merely takes photos from some sort of echo chamber, or house of mirrors
-- partly reflecting what is in it, and partly reproducing the observer (photographer)
in almost endless reflections.
Stitching together photos from these reflections yields a panorama of one and the same object in seemingly endless varieties.
In this way one might end up with a horribly distorted image of
this situation; and with the inside turned outside.

This is not dissimilar to what Plato outlined in the {\em Republic's} {\em cave metaphor}~\cite[Book~7, 515c, p.~221]{plato-republic}:
{\em ``what people in this situation would take for truth would be nothing more than the shadows of the manufactured objects.''}
In the quantum transcription of this metaphor,
the vectors are the objects, and the shadows taken for truth are the views on these objects,
mediated or translated~\cite{svozil-2003-garda} by arbitrary mismatching contexts.

\begin{acknowledgments}
Federico Holik has constantly inspired me to (re-)think probabilities and invited me to Argentina.
I kindly thank Ad\'an Cabello and Jos\'{e} R. Portillo for numerous explanations and discussions during an ongoing collaboration.
I am also deeply indebted to Alastair A. Abbott and Cristian S. Calude for their contributions and collaboration on the localization of value indefiniteness.
Josef Tkadlec has kindly provided a {\it Pascal} program to compute two-valued states on logics, given the contexts (blocks).
Christoph Lemell and the Nonlinear Dynamics group at the Vienna University of Technology
has provided  the computational framework for the Hull calculations performed.
\end{acknowledgments}


\appendix
\section{Two-valued states, (quasi)classical probabilities on the triangular logic in 4 dimensions}
\label{2018-whycontexts-app-tri}

The two-valued states ({\it aka} truth tables) have been enumerated by Josef Tkadlec's Pascal program {\tt 2states}~\cite{tkadlec-2states-2017}.
Implicitly, the convex sums over the respective probabilities
encode the truth tables, as, on any particular atom, the $i$'th truth table entry is 1 if $\lambda_i$ appears
in the listing of the classical probability $p_i$.
Otherwise, the $i$'th truth table entry is zero.

The bounds for classical probabilities have been obtained by Komei Fukuda's {\tt cddlib} package~\cite{cdd-pck}.


%


There are 9 propositions forming three contexts
$\{1,  2,  3,  4\}$,
$\{4,  5,  6,  7\}$, and
$\{7,  8,  9,  1\}$ allowing
14 (separating, unital) two-valued states whose convex sum yield the following (quasi)classical probabilities:
\begin{equation}
\begin{aligned}
p_{ 1} =&\lambda_1 + \lambda_2                                         ,\\
p_{ 2} =&\lambda_3 + \lambda_4 + \lambda_5 + \lambda_6 + \lambda_7           ,\\
p_{ 3} =&\lambda_8 + \lambda_9 + \lambda_{10}  + \lambda_{11}+ \lambda_{12}                                     ,\\
p_{ 4} =&\lambda_{13} + \lambda_{14}           ,\\
p_{ 5} =&\lambda_1 + \lambda_3 + \lambda_4 +\lambda_8 + \lambda_9                                         ,\\
p_{ 6} =&\lambda_2 + \lambda_5 + \lambda_6 + \lambda_{10} + \lambda_{11}          ,\\
p_{ 7} =&\lambda_7 + \lambda_{12}              ,\\
p_{ 8} =&\lambda_3 + \lambda_5 + \lambda_8 + \lambda_{10} + \lambda_{13}                                        ,\\
p_{ 9} =&\lambda_4 + \lambda_6 + \lambda_{9} + \lambda_{11} + \lambda_{14} .
\end{aligned}
\end{equation}

%


\section{Truth assignments, (quasi)classical probabilities  on the square logic in 4 dimensions}
\label{2018-whycontexts-app-square}

There are 12 propositions forming four contexts
$\{1,  2,  3,  4\}$,
$\{4,  5,  6,  7\}$,
$\{7,  8,  9,  10\}$,  and
$\{10,  11,  12,  1\}$
allowing  34 (separating, unital) two-valued states whose convex sum yield the following (quasi)classical probabilities:
\begin{equation}
\begin{aligned}
p_{ 1} =&\lambda_1 + \lambda_2+ \lambda_3+ \lambda_4+ \lambda_5,\\
p_{ 2} =&\lambda_6 + \lambda_7 + \lambda_8+ \lambda_9 + \lambda_{10}+ \lambda_{11}  \\& + \lambda_{12}+ \lambda_{13}+ \lambda_{14}+ \lambda_{15}+ \lambda_{16}+ \lambda_{17},\\
p_{ 3} =&\lambda_{18}+ \lambda_{19}+ \lambda_{20}+ \lambda_{21}+ \lambda_{22}+ \lambda_{23}  \\&+ \lambda_{24}+ \lambda_{25}+ \lambda_{26}+ \lambda_{24}+ \lambda_{28}+ \lambda_{29},\\
p_{ 4} =&\lambda_{30}+ \lambda_{31}+ \lambda_{32}+ \lambda_{33}+ \lambda_{34},\\
p_{ 5} =&\lambda_{1}+ \lambda_{2}+ \lambda_{6}+ \lambda_{7}+ \lambda_{8}+ \lambda_{9}  \\& + \lambda_{10}+ \lambda_{18}+ \lambda_{19}+ \lambda_{20}+ \lambda_{21}+ \lambda_{22},\\
p_{ 6} =&\lambda_{3}+ \lambda_{4}+ \lambda_{11}+ \lambda_{12}+ \lambda_{13}+ \lambda_{14}  \\& + \lambda_{15}+ \lambda_{23}+ \lambda_{24}+ \lambda_{25}+ \lambda_{26}+ \lambda_{27},\\
p_{ 7} =&\lambda_{5}+ \lambda_{16}+ \lambda_{17}+ \lambda_{28}+ \lambda_{29},\\
p_{ 8} =&\lambda_{1}+ \lambda_{3}+ \lambda_{6}+ \lambda_{7}+ \lambda_{11}+ \lambda_{12}  \\&+ \lambda_{18}+ \lambda_{19}+ \lambda_{23}+ \lambda_{24}+ \lambda_{30}+ \lambda_{31},\\
p_{ 9} =&\lambda_{2}+ \lambda_{4}+ \lambda_{8}+ \lambda_{9}+ \lambda_{13}+ \lambda_{14}  \\&+ \lambda_{20}+ \lambda_{21}+ \lambda_{25}+ \lambda_{26}+ \lambda_{32}+ \lambda_{33},\\
p_{10} =&\lambda_{10}+ \lambda_{15}+ \lambda_{22}+ \lambda_{27}+ \lambda_{34},\\
p_{11} =&\lambda_{6}+ \lambda_{8}+ \lambda_{11}+ \lambda_{13}+ \lambda_{16}+ \lambda_{18}  \\&+ \lambda_{20}+ \lambda_{23}+ \lambda_{25}+ \lambda_{28}+ \lambda_{30}+ \lambda_{32},\\
p_{12} =&\lambda_{7}+ \lambda_{9}+ \lambda_{12}+ \lambda_{14}+ \lambda_{17}+ \lambda_{19}  \\&+ \lambda_{21}+ \lambda_{24}+ \lambda_{26}+ \lambda_{29}+ \lambda_{31}+ \lambda_{33}.
\end{aligned}
\end{equation}

%

\section{Two-valued states, (quasi)classical probabilities  on the pentagon (pentagram) logic in 3 dimensions}
\label{2018-whycontexts-app-pent}


There are 5 contexts
$\{1,  2,  3\}$,
$\{3,  4,  5\}$,
$\{5,  6,  7\}$,
$\{7,  8,  9\}$,  and
$\{9,  10, 1\}$
allowing  11 (separating, unital) two-valued states~\cite{wright:pent} whose convex sum yield the following (quasi)classical probabilities:
\begin{equation}
\begin{aligned}
p_1=&\lambda_1 + \lambda_2 + \lambda_3 ,\\
p_2=&\lambda_4 + \lambda_5 + \lambda_7 + \lambda_9 + \lambda_{11}    ,\\
p_3=&\lambda_6 + \lambda_8 + \lambda_{10}                              ,\\
p_4=&\lambda_1 + \lambda_2 + \lambda_4 + \lambda_7 + \lambda_{11}      ,\\
p_5=&\lambda_3 + \lambda_5 + \lambda_9                                 ,\\
p_6=&\lambda_1 + \lambda_4 + \lambda_6 + \lambda_{10} + \lambda_{11}   ,\\
p_7=&\lambda_2 + \lambda_7 + \lambda_8                                 ,\\
p_8=&\lambda_1 + \lambda_3 + \lambda_9 + \lambda_{10} + \lambda_{11}   ,\\
p_9=&\lambda_4 + \lambda_5 + \lambda_6                                 ,\\
p_{10}=&\lambda_7 + \lambda_8 + \lambda_9 + \lambda_{10} + \lambda_{11}
.
\end{aligned}
\label{2018-whycontexts-ppent}
\end{equation}

%

\section{Truth assignments, (quasi)classical probabilities  on the Specker bug combo logic}
\label{2018-whycontexts-app-sbcombo}

The logic depicted in Fig.~\ref{2018-whycontexts-fbugcombo}(c)
contains 27 propositions forming 16 contexts
$\{ {\bf a},1,2           \}$,
$\{ 2,3,4           \}$,
$\{ 4,5,{\bf b}          \}$,
$\{ {\bf b},6,7           \}$,
$\{ 7,8,9           \}$,
$\{ 9,10,{\bf a}          \}$,
$\{ 3 ,  8  ,  11           \}$,
{\color{brown}$\{ {\bf a},{\bf c},{\bf b'}          \}$},
{\color{violet}$\{ {\bf b},{\bf c},{\bf a}'          \}$},
$\{ {\bf a}',1',2'        \}$,
$\{ 2',3',4'        \}$,
$\{ 4',5',{\bf b}'        \}$,
$\{ {\bf b}',6',7'        \}$,
$\{ 7',8',9'        \}$,
$\{ 9',10',{\bf a}'       \}$,   and
$\{ 3' ,  8'  ,  11'           \}$,
allowing  82 non-separating on ${\bf a}$/${\bf a'}$ and  ${\bf b}$/${\bf b'}$, unital two-valued states (not enumerated here because of volume).
9 and 9 of these permit ${\bf a}$ as well as ${\bf a'}$, and  ${\bf b}$ as well as ${\bf b'}$ to be true, respectively.

The logic depicted in Fig.~\ref{2018-whycontexts-fbugcombo}(b)
contains 16 propositions forming 9 contexts
$\{ {\bf a},1,2           \}$,
$\{ 2,3,4           \}$,
$\{ 4,5,{\bf b}          \}$,
$\{ {\bf b},6,7           \}$,
$\{ 7,8,9           \}$,
$\{ 9,10,{\bf a}          \}$,
$\{ 3 ,  8  ,  11           \}$,
{\color{brown}$\{ {\bf a},{\bf c},{\bf b'}          \}$},
{\color{violet}$\{ {\bf b},{\bf c},{\bf a}'          \}$},
allowing  22 (separating  and   unital) two-valued states which,
through their convex summation, yield the (quasi-)classical probabilities:
\begin{equation}
\begin{aligned}
 p_{{\bf a}}=&\lambda_1+\lambda_2+\lambda_3    ,\\
 p_{{\bf b}}=&\lambda_8+\lambda_{21}+\lambda_{22}  ,\\
 p_{{\bf a}'}=&\lambda_1+\lambda_2+\lambda_3+\lambda_5+\lambda_7+\lambda_{10}   \\&+\lambda_{12}+\lambda_{14}+\lambda_{16}+\lambda_{18}+\lambda_{20}  ,\\
 p_{{\bf b}'}=&\lambda_5+\lambda_7+\lambda_8+\lambda_{10}+\lambda_{12}+\lambda_{14} \\& +\lambda_{16}+\lambda_{18}+\lambda_{20}+\lambda_{21}+\lambda_{22}  , \\
 p_{{\bf c}}=&\lambda_4+\lambda_6+\lambda_9+\lambda_{11}+\lambda_{13}+\lambda_{15}+\lambda_{17}+\lambda_{19} ,\\
 p_{ 1}=&\lambda_4+\lambda_5+\lambda_6+\lambda_7+\lambda_8+\lambda_9  \\& +\lambda_{10}+\lambda_{11}+\lambda_{12}+\lambda_{13}+\lambda_{14}  ,\\
 p_{ 2}=&\lambda_{15}+\lambda_{16}+\lambda_{17}+\lambda_{18}+\lambda_{19}+\lambda_{20}+\lambda_{21}+\lambda_{22}  ,\\
 p_{ 3}=&\lambda_1+\lambda_4+\lambda_5+\lambda_6+\lambda_7+\lambda_8      ,\\
 p_{ 4}=&\lambda_2+\lambda_3+\lambda_9+\lambda_{10}+\lambda_{11} +\lambda_{12}+\lambda_{13}+\lambda_{14}  ,\\
 p_{ 5}=&\lambda_1+\lambda_4+\lambda_5+\lambda_6+\lambda_7+\lambda_{15} \\& +\lambda_{16}+\lambda_{17}+\lambda_{18}+\lambda_{19}+\lambda_{20}  ,\\
 p_{ 6}=&\lambda_2+\lambda_4+\lambda_5+\lambda_9+\lambda_{10}+\lambda_{11} \\& +\lambda_{12}+\lambda_{15}+\lambda_{16}+\lambda_{17}+\lambda_{18}  ,\\
 p_{ 7}=&\lambda_1+\lambda_3+\lambda_6+\lambda_7+\lambda_{13}+\lambda_{14}+\lambda_{19}+\lambda_{20}  ,\\
 p_{ 8}=&\lambda_2+\lambda_9+\lambda_{10}+\lambda_{15}+\lambda_{16}+\lambda_{21}  ,\\
 p_{ 9}=&\lambda_4+\lambda_5+\lambda_8+\lambda_{11}+\lambda_{12} +\lambda_{17}+\lambda_{18}+\lambda_{22}  ,\\
 p_{10}=&\lambda_6+\lambda_7+\lambda_9+\lambda_{10}+\lambda_{13} +\lambda_{14} \\& +\lambda_{15}+\lambda_{16}+\lambda_{19}+\lambda_{20}+\lambda_{21}  ,\\
 p_{11}=&\lambda_3+\lambda_{11}+\lambda_{12}+\lambda_{13}+\lambda_{14} +\lambda_{17} \\& +\lambda_{18}+\lambda_{19}+\lambda_{20}+\lambda_{22}
.
\end{aligned}
\end{equation}
Note that, for all configurations, $p_{{\bf a}} =\lambda_1+\lambda_2+\lambda_3  \le  p_{{\bf a}'}$,
implying that, whenever ${\bf a}$ is true  ${\bf a}'$ must be true as well.

The Specker bug logic depicted in Fig.~\ref{2018-whycontexts-fbugcombo}(a)
contains 13 propositions forming 7 contexts
$\{ {\bf a},1,2           \}$,
$\{ 2,3,4           \}$,
$\{ 4,5,{\bf b}          \}$,
$\{ {\bf b},6,7           \}$,
$\{ 7,8,9           \}$,
$\{ 9,10,{\bf a}          \}$,
$\{ 3 ,  8  ,  11           \}$,
allowing  14 (separating  and   unital) two-valued states:
\begin{equation}
\begin{aligned}
 p_{{\bf a}}=&\lambda_1+\lambda_2+\lambda_3          ,\\
 p_{{\bf b}}=&\lambda_6+\lambda_{13}+\lambda_{14}   ,\\
 p_{ 1}=&\lambda_4+\lambda_5+\lambda_6+\lambda_7+\lambda_8+\lambda_9     ,\\
 p_{ 2}=&\lambda_{10}+\lambda_{11}+\lambda_{12}+\lambda_{13}+\lambda_{14}  ,\\
 p_{ 3}=&\lambda_1+\lambda_4+\lambda_5+\lambda_6        ,\\
 p_{ 4}=&\lambda_2+\lambda_3+\lambda_7+\lambda_8+\lambda_9      ,\\
 p_{ 5}=&\lambda_1+\lambda_4+\lambda_5+\lambda_{10}+\lambda_{11}+\lambda_{12}  ,\\
 p_{ 6}=&\lambda_2+\lambda_4+\lambda_7+\lambda_8+\lambda_{10}+\lambda_{11}  ,\\
 p_{ 7}=&\lambda_1+\lambda_3+\lambda_5+\lambda_9+\lambda_{12}  ,\\
 p_{ 8}=&\lambda_2+\lambda_7+\lambda_{10}+\lambda_{13}  ,\\
 p_{ 9}=&\lambda_4+\lambda_6+\lambda_8+\lambda_{11}+\lambda_{14}  ,\\
 p_{10}=&\lambda_5+\lambda_7+\lambda_9+\lambda_{10}+\lambda_{12}+\lambda_{13}  ,\\
 p_{11}=&\lambda_3+\lambda_8+\lambda_9+\lambda_{11}+\lambda_{12}+\lambda_{14}
.
\end{aligned}
\label{2018-whycontexts-pspeckerbug}
\end{equation}
Note that, for all configurations, whenever ${\bf a}$ is true {\bf b} is false, and {\it vice versa}.

\section{Truth assignments, (quasi)classical probabilities  on truth-implies-value indefiniteness logic in 3 dimensions}
\label{2018-whycontexts-app-ACS}

Fig.~\ref{2018-whycontexts-impossible}(c)
depicts 37 propositions $\{{\bf a}, {\bf b}, 1, 2, 3, \ldots , 35\}$ in 26 contexts
$\{ {\bf a}  ,1   ,2       \}$,
$\{ {\bf b}  ,2   ,3       \}$,
$\{ 4  ,{\bf a}   ,5       \}$,
$\{ {\bf b}  ,6   ,7       \}$,
{\color{magenta}$[\{ 7  ,10  , 4      \}]_\textrm{(a),(c)}$},
{\color{magenta}$[\{ 10 , 12 ,  13    \}]_\textrm{(a),(c)}$},
{\color{blue}$[\{ 5  ,29  , 23     \}]_\textrm{(b),(c)}$},
{\color{blue}$[\{ 13 , 31 ,  29    \}]_\textrm{(b),(c)}$},
$\{ 3  ,21  , 23     \}$,
$\{ 4  ,28  , 22     \}$,
$\{ 22 , 19 ,  3     \}$,
$\{ {\bf b}  ,8   ,9       \}$,
$\{ 9  ,11  , 5      \}$,
$\{ 28 , 30 ,  15    \}$,
$\{ 15 , 14 ,  11    \}$,
$\{ 6  ,33  , 17     \}$,
$\{ 17 , 20 ,  21    \}$,
$\{ 7  ,34  , 27     \}$,
$\{ 27 , 26 ,  23    \}$,
$\{ 22 , 24 ,  25    \}$,
$\{ 25 , 35 ,  9     \}$,
$\{ 15 , 17 ,  1     \}$,
$\{ 13 , 16 ,  1     \}$,
$\{ 16 , 18 ,  19    \}$,
$\{ 16 , 32 ,  8     \}$,
$\{ 25 , 1  , 27     \}$,
allowing  8 (non-separating, non-unital on ${\bf a}$, 2, 13, 15, 16, 17, 25, 27)
two-valued states whose convex sum yield the following weights:
\begin{equation}
\begin{aligned}
p_{{\bf a}}=&p_{ 2}=  p_{13}= p_{15}= p_{16}=p_{17}=p_{25}= p_{27}=  0                                                                                       , \\
 p_{{\bf b}}=&            \lambda_1   +\lambda_2   +\lambda_3   +\lambda_4  ,\\
 p_{ 1}=&           \lambda_1   +\lambda_2   +\lambda_3   +\lambda_4   +\lambda_5   +\lambda_6   +\lambda_7   +\lambda_8 = 1                , \\
 p_{ 3}=&                                                                   +\lambda_5   +\lambda_6   +\lambda_7   +\lambda_8               , \\
 p_{ 4}=&           \lambda_1   +\lambda_2                               +\lambda_5   +\lambda_6                                            , \\
 p_{ 5}=&                                       \lambda_3   +\lambda_4                               +\lambda_7   +\lambda_8               , \\
 p_{ 6}=&                                                                   \lambda_5   +\lambda_6   +\lambda_7                            , \\
 p_{ 7}=&\lambda_8           ,  p_{ 9}=   \lambda_6   ,  p_{22}= \lambda_4,  p_{23}=                         \lambda_2  ,  \\
 p_{ 8}=&                                                                   \lambda_5                 +\lambda_7   +\lambda_8              , \\
 p_{10}=&                                       \lambda_3   +\lambda_4                               +\lambda_7                            , \\
 p_{11}=&           \lambda_1   +\lambda_2                               +\lambda_5                                                         , \\
 p_{12}=&           \lambda_1   +\lambda_2                               +\lambda_5   +\lambda_6                 +\lambda_8                 , \\
 p_{14}=&                                       \lambda_3   +\lambda_4                 +\lambda_6   +\lambda_7   +\lambda_8                , \\
 p_{18}=&                                                    \lambda_4   +\lambda_5   +\lambda_6   +\lambda_7   +\lambda_8                , \\
 p_{19}=&           \lambda_1   +\lambda_2   +\lambda_3                                                                                     , \\
 p_{20}=&                         +\lambda_2                               +\lambda_5   +\lambda_6   +\lambda_7   +\lambda_8                , \\
 p_{21}=&           \lambda_1                 +\lambda_3   +\lambda_4                                                                       , \\
 p_{24}=&           \lambda_1   +\lambda_2   +\lambda_3                 +\lambda_5   +\lambda_6   +\lambda_7   +\lambda_8                   , \\
 p_{26}=&           \lambda_1                 +\lambda_3   +\lambda_4   +\lambda_5   +\lambda_6   +\lambda_7   +\lambda_8                   , \\
 p_{28}=&                                       \lambda_3                                             +\lambda_7   +\lambda_8              , \\
 p_{29}=&           \lambda_1     +                                        \lambda_5   +\lambda_6                                           , \\
 p_{30}=&           \lambda_1   +\lambda_2                 +\lambda_4   +\lambda_5   +\lambda_6                                             , \\
 p_{31}=&                         \lambda_2   +\lambda_3   +\lambda_4                               +\lambda_7   +\lambda_8                , \\
 p_{32}=&           \lambda_1   +\lambda_2   +\lambda_3   +\lambda_4                 +\lambda_6                                             , \\
 p_{33}=&           \lambda_1   +\lambda_2   +\lambda_3   +\lambda_4                                             +\lambda_8                 , \\
 p_{34}=&           \lambda_1   +\lambda_2   +\lambda_3   +\lambda_4   +\lambda_5   +\lambda_6   +\lambda_7                                 , \\
 p_{35}=&           \lambda_1   +\lambda_2   +\lambda_3   +\lambda_4   +\lambda_5                 +\lambda_7   +\lambda_8
.
\end{aligned}
\end{equation}

The logics in
Figs.~\ref{2018-whycontexts-impossible}(a)
and~\ref{2018-whycontexts-impossible}(b)
contain 35 observables in 24 contexts which are the same as before in Fig.~\ref{2018-whycontexts-impossible}(c)
lacking two contexts
{\color{blue}$[\{ 5  ,29  , 23     \}]_\textrm{(b),(c)}$} and
{\color{blue}$[\{ 13 , 31 ,  29    \}]_\textrm{(b),(c)}$},
as well as
{\color{magenta}$[\{ 7  ,10  , 4      \}]_\textrm{(a),(c)}$} and
{\color{magenta}$[\{ 10 , 12 ,  13    \}]_\textrm{(a),(c)}$},
respectively.

The logic in
Figs.~\ref{2018-whycontexts-impossible}(a)
allows  13 (non-unital on  16)
two-valued states whose convex sum yield the following weights:
\begin{equation}
\begin{aligned}
 p_{{\bf a}}=&\lambda_1 ,\\
 p_{{\bf b}}=&\lambda_2+\lambda_3+\lambda_4+\lambda_5+\lambda_6+\lambda_7 ,\\
 p_{16}=& 0,\\
 p_{ 1}=&\lambda_2+\lambda_3+\lambda_4+\lambda_5+\lambda_6+\lambda_7+\lambda_8+\lambda_9+\lambda_{10}+\lambda_{11}  ,\\
 p_{ 2}=&\lambda_{12}+\lambda_{13}  ,\\
 p_{ 3}=&\lambda_1+\lambda_8+\lambda_9+\lambda_{10}+\lambda_{11}  ,\\
 p_{ 4}=&\lambda_2+\lambda_3+\lambda_8+\lambda_9+\lambda_{12}  ,\\
 p_{ 5}=&\lambda_4+\lambda_5+\lambda_6+\lambda_7+\lambda_{10}+\lambda_{11}+\lambda_{13}  ,\\
 p_{ 6}=&\lambda_8+\lambda_9+\lambda_{10}+\lambda_{12}  ,\\
 p_{ 7}=&\lambda_1+\lambda_{11}+\lambda_{13}  ,\\
 p_{ 8}=&\lambda_1+\lambda_8+\lambda_{10}+\lambda_{11}+\lambda_{13}  ,\\
 p_{ 9}=&\lambda_9+\lambda_{12}  ,\\
 p_{10}=&\lambda_4+\lambda_5+\lambda_6+\lambda_7+\lambda_{10}  ,\\
 p_{11}=&\lambda_1+\lambda_2+\lambda_3+\lambda_8 ,\\
 p_{12}=&\lambda_2+\lambda_3+\lambda_8+\lambda_9+\lambda_{11}  ,\\
 p_{13}=&\lambda_1+\lambda_{12}+\lambda_{13}  ,\\
 p_{14}=&\lambda_4+\lambda_5+\lambda_6+\lambda_7+\lambda_9+\lambda_{10}+\lambda_{11}+\lambda_{13}  ,\\
 p_{15}=&\lambda_{12}  ,\\
 p_{17}=&\lambda_1+\lambda_{13}  ,\\
 p_{18}=&\lambda_1+\lambda_5+\lambda_7+\lambda_8+\lambda_9+\lambda_{10}+\lambda_{11}  ,\\
 p_{19}=&\lambda_2+\lambda_3+\lambda_4+\lambda_6+\lambda_{12}+\lambda_{13}  ,\\
 p_{20}=&\lambda_3+\lambda_6+\lambda_7+\lambda_8+\lambda_9+\lambda_{10}+\lambda_{11}  ,\\
 p_{21}=&\lambda_2+\lambda_4+\lambda_5+\lambda_{12}  ,\\
 p_{22}=&\lambda_5+\lambda_7  ,\\
 p_{23}=&\lambda_3+\lambda_6+\lambda_7+\lambda_{13}  ,\\
 p_{24}=&\lambda_2+\lambda_3+\lambda_4+\lambda_6+\lambda_8+\lambda_9+\lambda_{10}+\lambda_{11}+\lambda_{12}  ,\\
 p_{25}=&\lambda_1+\lambda_{13}  ,\\
 p_{26}=&\lambda_1+\lambda_2+\lambda_4+\lambda_5+\lambda_8+\lambda_9+\lambda_{10}+\lambda_{11}  ,\\
 p_{27}=&\lambda_{12}  ,\\
 p_{28}=&\lambda_1+\lambda_4+\lambda_6+\lambda_{10}+\lambda_{11}+\lambda_{13}  ,\\
 p_{30}=&\lambda_2+\lambda_3+\lambda_5+\lambda_7+\lambda_8+\lambda_9 ,\\
 p_{32}=&\lambda_2+\lambda_3+\lambda_4+\lambda_5+\lambda_6+\lambda_7+\lambda_9+\lambda_{12}  ,\\
 p_{33}=&\lambda_2+\lambda_3+\lambda_4+\lambda_5+\lambda_6+\lambda_7+\lambda_{11}  ,\\
 p_{34}=&\lambda_2+\lambda_3+\lambda_4+\lambda_5+\lambda_6+\lambda_7+\lambda_8+\lambda_9+\lambda_{10}  ,\\
 p_{35}=&\lambda_2+\lambda_3+\lambda_4+\lambda_5+\lambda_6+\lambda_7+\lambda_8+\lambda_{10}+\lambda_{11}
.
\end{aligned}
\end{equation}
Therefore, whenever ${\bf a}$ is true, that is,
$p_{{\bf a}}=\lambda_1 =1$,
${\bf b}$ has to be false, because
$p_{{\bf b}}=\lambda_2+\lambda_3+\lambda_4+\lambda_5+\lambda_6+\lambda_7 =0$.

Conversely,
the logic in
Figs.~\ref{2018-whycontexts-impossible}(b)
allows  13 (non-separating on 15/27 and non-unital on 16)
two-valued states whose convex sum yield the following weights:
\begin{equation}
\begin{aligned}
 p_{{\bf a}}=&\lambda_1,\\
 p_{{\bf b}}=&\lambda_1+\lambda_2+\lambda_3+\lambda_4+\lambda_5,\\
 p_{16}=& 0,\\
 p_{ 1}=&\lambda_2+\lambda_3+\lambda_4+\lambda_5+\lambda_6+\lambda_7+\lambda_8+\lambda_9+\lambda_{10}+\lambda_{11}  ,\\
 p_{ 2}=&\lambda_{12}+\lambda_{13}  ,\\
 p_{ 3}=&\lambda_6+\lambda_7+\lambda_8+\lambda_9+\lambda_{10}+\lambda_{11}  ,\\
 p_{ 4}=&\lambda_2+\lambda_3+\lambda_6+\lambda_7+\lambda_8+\lambda_9+\lambda_{12}  ,\\
 p_{ 5}=&\lambda_4+\lambda_5+\lambda_{10}+\lambda_{11}+\lambda_{13}  ,\\
 p_{ 6}=&\lambda_6+\lambda_7+\lambda_{10}+\lambda_{13}  ,\\
 p_{ 7}=&\lambda_8+\lambda_9+\lambda_{11}+\lambda_{12}  ,\\
 p_{ 8}=&\lambda_6+\lambda_8+\lambda_{10}+\lambda_{11}+\lambda_{12}+\lambda_{13}  ,\\
 p_{ 9}=&\lambda_7+\lambda_9,\\
 p_{11}=&\lambda_1+\lambda_2+\lambda_3+\lambda_6+\lambda_8+\lambda_{12}  ,\\
 p_{13}=&\lambda_1+\lambda_{12}+\lambda_{13}  ,\\
 p_{14}=&\lambda_4+\lambda_5+\lambda_7+\lambda_9+\lambda_{10}+\lambda_{11}  ,\\
 p_{15}=& p_{27}=\lambda_{13}  ,\\
 p_{17}=&\lambda_1+\lambda_{12}  ,\\
 p_{18}=&\lambda_5+\lambda_6+\lambda_7+\lambda_8+\lambda_9+\lambda_{10}+\lambda_{11}+\lambda_{13}  ,\\
 p_{19}=&\lambda_1+\lambda_2+\lambda_3+\lambda_4+\lambda_{12}  ,\\
 p_{20}=&\lambda_3+\lambda_6+\lambda_7+\lambda_8+\lambda_9+\lambda_{10}+\lambda_{11}  ,\\
 p_{21}=&\lambda_2+\lambda_4+\lambda_5+\lambda_{13}  ,\\
 p_{22}=&\lambda_5+\lambda_{13}  ,\\
 p_{23}=&\lambda_1+\lambda_3+\lambda_{12}  ,\\
 p_{24}=&\lambda_2+\lambda_3+\lambda_4+\lambda_6+\lambda_7+\lambda_8+\lambda_9+\lambda_{10}+\lambda_{11}  ,\\
 p_{25}=&\lambda_1+\lambda_{12}  ,\\
 p_{26}=&\lambda_2+\lambda_4+\lambda_5+\lambda_6+\lambda_7+\lambda_8+\lambda_9+\lambda_{10}+\lambda_{11}  ,\\
 p_{28}=&\lambda_1+\lambda_4+\lambda_{10}+\lambda_{11}  ,\\
 p_{29}=&\lambda_2+\lambda_6+\lambda_7+\lambda_8+\lambda_9 ,\\
 p_{30}=&\lambda_2+\lambda_3+\lambda_5+\lambda_6+\lambda_7+\lambda_8+\lambda_9+\lambda_{12}  ,\\
 p_{31}=&\lambda_3+\lambda_4+\lambda_5+\lambda_{10}+\lambda_{11}  ,\\
 p_{32}=&\lambda_1+\lambda_2+\lambda_3+\lambda_4+\lambda_5+\lambda_7+\lambda_9,\\
 p_{33}=&\lambda_2+\lambda_3+\lambda_4+\lambda_5+\lambda_8+\lambda_9+\lambda_{11}  ,\\
 p_{34}=&\lambda_1+\lambda_2+\lambda_3+\lambda_4+\lambda_5+\lambda_6+\lambda_7+\lambda_{10}  ,\\
 p_{35}=&\lambda_2+\lambda_3+\lambda_4+\lambda_5+\lambda_6+\lambda_8+\lambda_{10}+\lambda_{11}+\lambda_{13}
.
\end{aligned}
\end{equation}
Therefore, whenever ${\bf a}$ is true, that is,
$p_{{\bf a}}=\lambda_1 =1$,
${\bf b}$ has to be true, because
$p_{{\bf b}}=\lambda_1+\lambda_2+\lambda_3+\lambda_4+\lambda_5 =\lambda_1=1$.

\clearpage



%

\end{document}